%% file: main.tex
\documentclass[%
letterpaper]{report}

\usepackage[margin=1in]{geometry}

\usepackage[hyphens]{url}

\usepackage[backend=biber,
style=verbose-inote,
]{biblatex}
\usepackage{etoolbox}
\patchcmd{\section}
  {\centering}
  {\raggedright}
  {}
  {}
\patchcmd{\subsection}
  {\centering}
  {\raggedright}
  {}
  {}
  
\counterwithout{footnote}{chapter}

\usepackage{titling}

\usepackage{clipboard}
\usepackage{times}
\usepackage{graphicx}
\usepackage{xcolor}
\usepackage{dcolumn}
\usepackage{soul}
\usepackage{bm}
\usepackage[normalem]{ulem} 
\usepackage{tabto}
\usepackage{amsmath}
\usepackage{amssymb}

\usepackage[pagewise]{lineno}

\definecolor{xlinkcolor}{cmyk}{1,1,0,0}
\usepackage{url}
\usepackage[
 colorlinks=true,    
 linkcolor=xlinkcolor,     
 citecolor=xlinkcolor,     
 filecolor=xlinkcolor,  
 urlcolor=xlinkcolor,      
 final=true
 pdftitle={Nu Tools: Exploring Practical Roles for Neutrinos in Nuclear Energy and Security}
 bookmarks=true
 pdfpagemode=FullScreen
]{hyperref}

\usepackage[T1]{fontenc}
\usepackage{titlesec, blindtext, color}
\definecolor{gray75}{gray}{0.75}
\newcommand{\hsp}{\hspace{20pt}}
\titleformat{\chapter}[hang]{\Huge\bfseries}{\thechapter\hsp\textcolor{gray75}{|}\hsp}{0pt}{\Huge\bfseries}

\usepackage{pdfpages}

\usepackage{comment}
\usepackage{xcolor}

\definecolor{nbblue}{rgb}{0.0, 0.58, 0.71}
\definecolor{phgreen}{rgb}{0.0, 0.42, 0.24}
\definecolor{acpurp}{rgb}{0.73, 0.33, 0.83}
\definecolor{tared}{rgb}{0.73, 0.33, 0}

\definecolor{execcolor}{rgb}{0.48, 0.25, 0}

\newcommand{\DNNRD}{Defense Nuclear Nonproliferation Research and Development (DNN R\&D)\renewcommand{\DNNRD}{DNN R\&D}}
\newcommand{\NRC}{Nuclear Regulatory Commission (NRC)\renewcommand{\NRC}{NRC}}
\newcommand{\TRL}{Technical Readiness Level (TRL)\renewcommand{\TRL}{TRL}}
\newcommand{\NPT}{Treaty on the Non-Proliferation of Nuclear Weapons (NPT)\renewcommand{\NPT}{NPT}}
\newcommand{\PMDA}{Plutonium Management and Disposition Agreement (PMDA)\renewcommand{\PMDA}{PMDA}}
\newcommand{\JCPOA}{Joint Comprehensive Plan of Action (JCPOA)\renewcommand{\JCPOA}{JCPOA}}
\newcommand{\FMCT}{Fissile Material Cutoff Treaty (FMCT)\renewcommand{\FMCT}{FMCT}}
\newcommand{\IAEA}{International Atomic Energy Agency (IAEA)\renewcommand{\IAEA}{IAEA}}
\newcommand{\BWR}{Boiling Water Reactors (BWR)\renewcommand{\BWR}{BWR}}

\newcommand{\CCFRefTechnicalReadiness}{\hyperref[finding:TechnicalReadiness]{\emph{Technical Readiness Finding}}} 

\newcommand{\CCFRefSiting}{\hyperref[finding:Siting]{\emph{Neutrino System Siting Finding}}} 

\newcommand{\UCFRefCurrentIAEA}{\hyperref[finding:CurrentIAEA]{\emph{Current \IAEA\ Safeguards Finding}}} 

\newcommand{\UCFRefAdvancedReactors}{\hyperref[finding:AdvancedReactors]{\emph{Advanced Reactors Finding}}}

\newcommand{\UCFRefReactorOperations}{\hyperref[finding:ReactorOperations]{\emph{Reactor Operations Finding}}}

\newcommand{\UCFRefPostAccident}{\hyperref[finding:PostAccident]{\emph{Post-Accident Response Finding}}} 

\newcommand{\UCFRefSpentFuel}{\hyperref[finding:SpentFuel]{\emph{Spent Nuclear Fuel Finding}}} 

\newcommand{\ChapRefCCFindings}{\hyperref[crossCuttingFindings]{\emph{Cross-Cutting Findings}}} 

\newcommand{\ChapRefUCFindings}{\hyperref[useCaseFindings]{\emph{Use Case Findings}}} 

\newcommand{\ChapRefFramework}{\hyperref[utilityFramework]{\emph{Framework for Evaluating Utility}}}

\usepackage[draft, noinline, nomargin]{fixme}
\fxsetup{theme=color, mode=multiuser}

\usepackage{fancyhdr}
\title{Nu Tools: Exploring Practical Roles for Neutrinos \\ in Nuclear Energy and Security}

\usepackage{bigfoot}

\DeclareNewFootnote{ANote}[fnsymbol]

\usepackage{etoolbox}
\makeatletter
\patchcmd\maketitle{\def\@makefnmark{\rlap{\@textsuperscript{\normalfont\@thefnmark}}}}{}{}{}
\makeatother

\makeatletter
\def\thanksANote#1{%
  \protected@xdef\@thanks{\@thanks%
        \protect\footnotetextANote[\the \c@footnoteANote]{#1}}%
}
\makeatother

\author{
  Final Report of the Antineutrino Reactor Monitoring Scoping Study
}
  \thanksANote{PNNL-31870, \href{https://doi.org/10.2172/1826602}{doi:10.2172/1826602}}%

\date{\today}

\begin{document}

\pagestyle{empty}

\includepdf[pages=-]{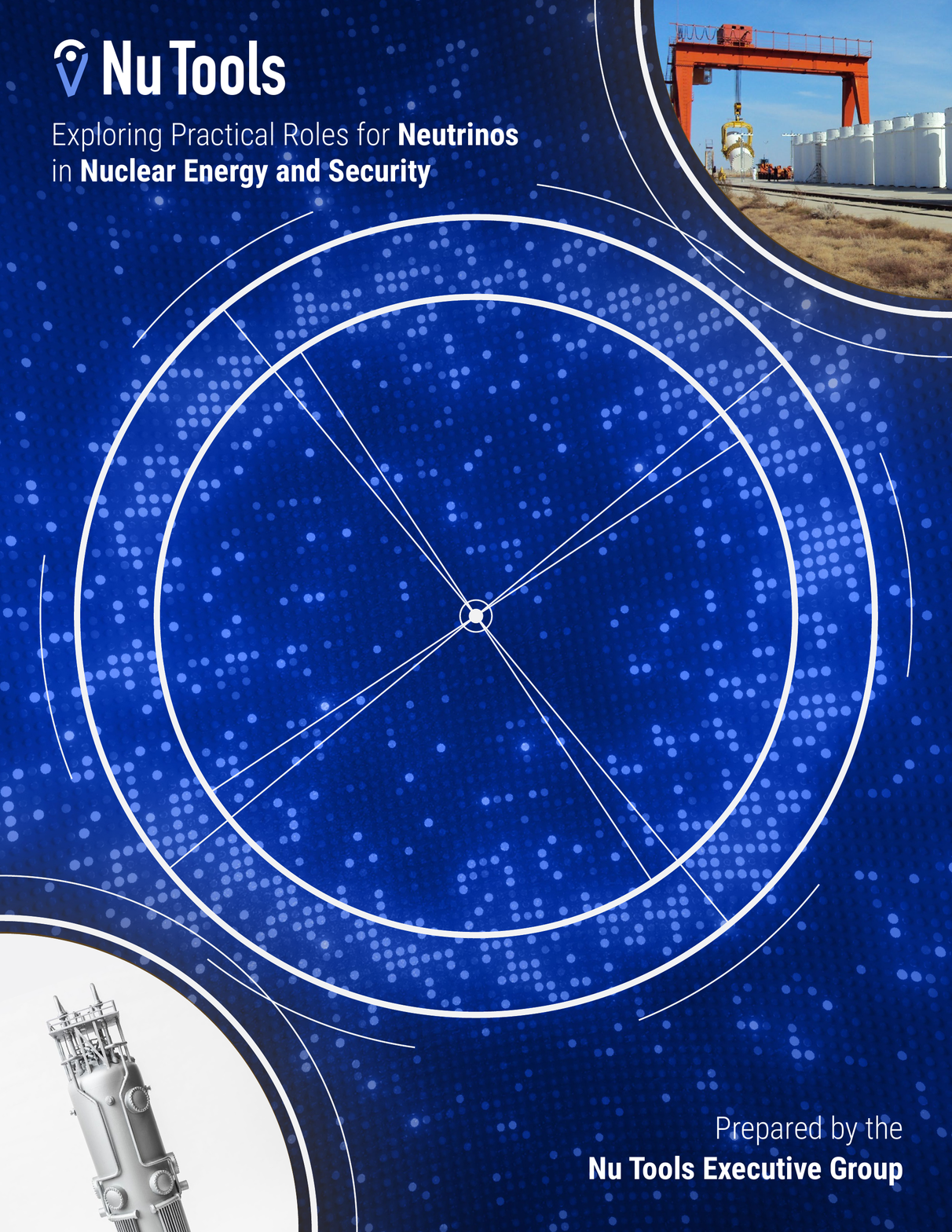}

\includepdf[pages=-]{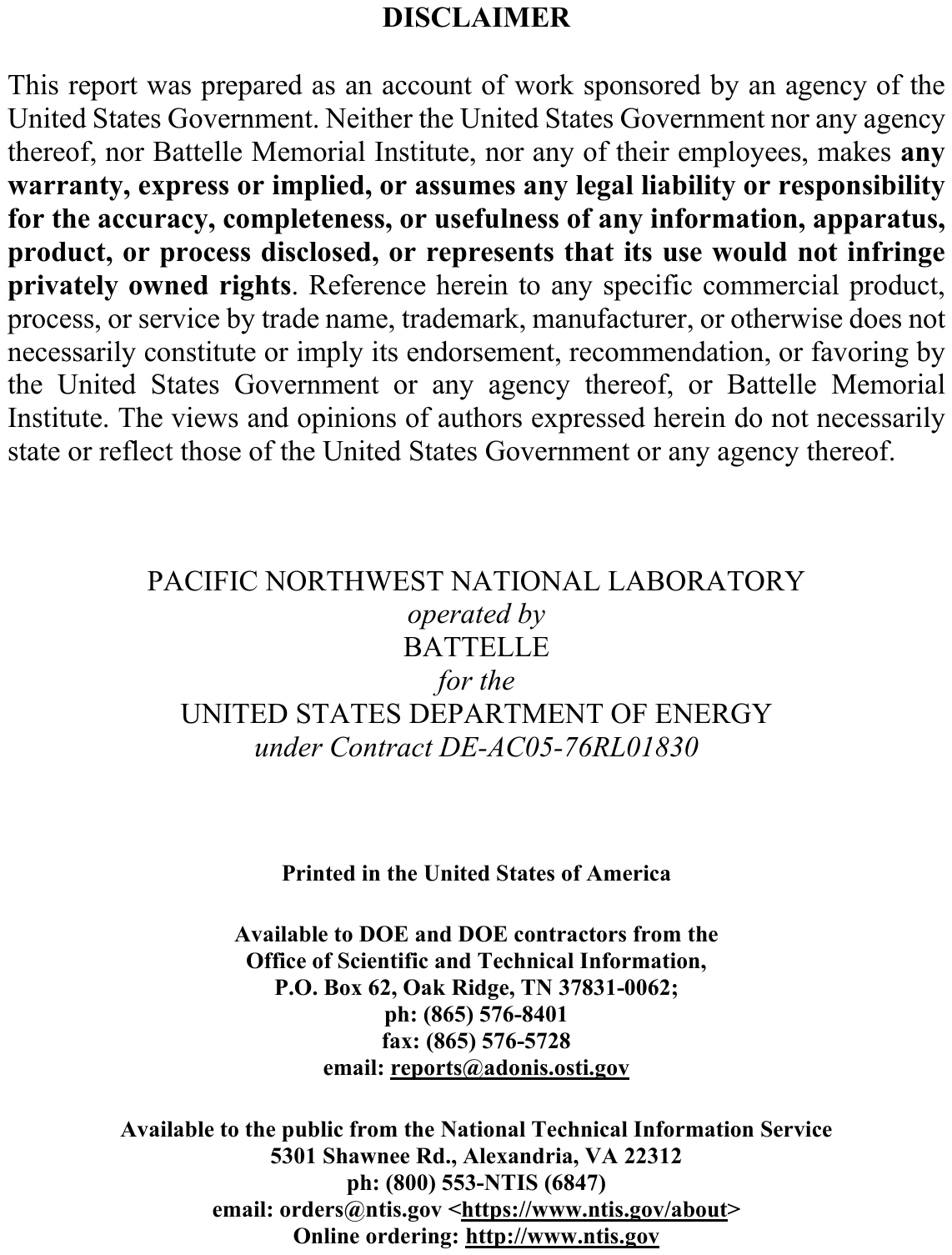}

\maketitle

\pagestyle{fancy}

\section*{Study Authors}

This report was prepared by the Nu Tools Executive group:

\begin{itemize}
\TabPositions{5cm} 
\item[] Oluwatomi Akindele \tab \textit{Lawrence Livermore National Laboratory}
\item[] Nathaniel Bowden \tab \textit{Lawrence Livermore National Laboratory}
\item[] Rachel Carr \tab \textit{Massachusetts Institute of Technology}
\item[] Andrew Conant \tab \textit{Oak Ridge National Laboratory}
\item[] Milind Diwan \tab \textit{Brookhaven National Laboratory} 
\item[] Anna Erickson \tab \textit{Georgia Institute of Technology}
\item[] Michael Foxe \tab \textit{Pacific Northwest National Laboratory} 
\item[] Bethany L. Goldblum \tab \textit{Lawrence Berkeley National Laboratory; University of California, Berkeley} 
\item[] Patrick Huber \tab \textit{Virginia Tech}
\item[] Igor Jovanovic \tab \textit{University of Michigan} 
\item[] Jonathan Link \tab \textit{Virginia Tech} 
\item[] Bryce Littlejohn \tab \textit{Illinois Institute of Technology} 
\item[] Pieter Mumm \tab \textit{National Institute of Standards and Technology}
\item[] Jason Newby \tab \textit{Oak Ridge National Laboratory}

\end{itemize}

\newpage

\section*{Acknowledgements}

\noindent A portion of this work was performed under the auspices of the U.S. Department of Energy by Lawrence Berkeley National Laboratory under Contract DE-AC02-05CH11231. The project was funded by the U.S. Department of Energy, National Nuclear Security Administration (NNSA), Office of Defense Nuclear Nonproliferation Research and Development (DNN R\&D).
\\

\noindent A portion of this work was performed under the auspices of the U.S. Department of Energy by Brookhaven National Laboratory under contract DE-SC0012704. The project was funded by the U.S. Department of Energy, Office of Science and National Nuclear Security Administration (NNSA), Office of Defense Nuclear Nonproliferation Research and Development (DNN R\&D).
\\

\noindent A portion of this work was performed under the auspices of the U.S. Department of Energy by Lawrence Livermore National Laboratory under Contract DE-AC52-07NA27344. This research was supported by the Office of Defense Nuclear Nonproliferation R\&D (DNN R\&D) in the National Nuclear Security Administration (NNSA). \\

\noindent A portion of this research was supported by the Office of Defense Nuclear Nonproliferation R\&D (DNN R\&D) in the National Nuclear Security Administration (NNSA), U.S. Department of Energy under contract DE-AC05-00OR22725 with Oak Ridge National Laboratory, managed and operated by UT-Battelle, LLC. \\

\noindent A portion of this work was performed by Pacific Northwest National Laboratory, a contractor of the U.S. Department of Energy, under contract DE-AC05-76RL01830. This research was supported by the Office of Defense Nuclear Nonproliferation R\&D (DNN R\&D) in the National Nuclear Security Administration (NNSA).\\

\noindent Portions of this work were supported by the U. S. Department of Energy National Nuclear Security Administration (NNSA) through the Nuclear Science and Security Consortium under award number DE-NA0003180, the Monitoring, Technology and Verification Consortium under award number DE-NA0003920, and the Consortium for Enabling Technologies and Innovation under award number DE-NA0003921, and by the National Science Foundation under award numbers IIP-1924433 and IIP-1941238. \\

\noindent Portions of this work were supported by the National Institute of Standards and Technology within the U.S. Department of Commerce.\\

\noindent National Laboratory Release Information:\\
\noindent BNL-222043-2021-JAAM \\
\noindent LLNL-TR-826137 \\
\noindent PNNL-31870 \\

\newpage

\tableofcontents

\newpage

\chapter{Executive Summary}

\input{execSummary}

\newpage

\chapter{Study Approach}
\label{studyapproach}

\input{studyIntro}

\newpage

\chapter{Cross-Cutting Findings}

\input{crossCuttingFindings}

\newpage

\chapter{Framework for Evaluating Utility}

\input{utilityFramework}

\newpage

\chapter{Use Case Findings}

\input{useCaseFindings}

\newpage

\chapter{Recommendations}

\input{recommendations}

\newpage

\appendix

\chapter{Glossary of Terms}
\input{glossary}

\chapter{ARMS Study Charge}
\label{charge}

\includepdf[pages=-,pagecommand={},width=1.15\textwidth]{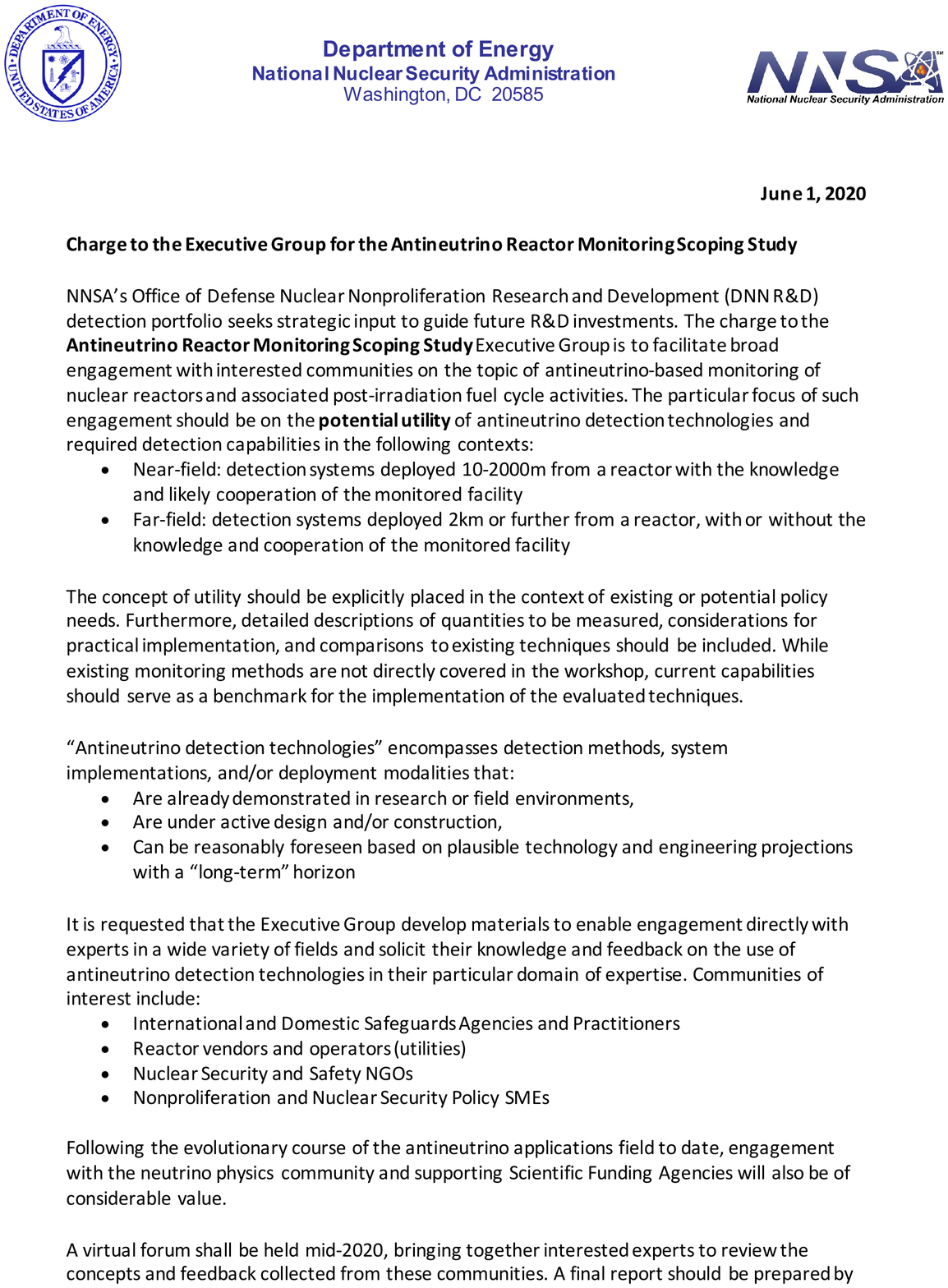}

\chapter{Interviewee List}
\label{ch:interviewee_list}
\input{interviewees}

\chapter{Synopsis of the Nu Tools Mini-Workshop for the Applied Antineutrino Technology Community}
\label{miniWorkshop}
\input{miniWorkshop}

\chapter{Fact Sheets for End-User Engagement}
\label{FactSheets}



\section*{Supplementary material}

# Quick Introduction to Neutrino Detectors

In a nuclear reactor, fission emits large numbers of subatomic particles called *neutrinos*. These particles leave the reactor building in all directions and cannot be shielded. Detection technology now exists to measure these emissions and potentially use them to monitor reactors and associated facilities. Reactor neutrino detection has been demonstrated at distances of 10 m to 100 km, aboveground and belowground, and with corresponding detector sizes of 1–1,000 metric tons.

**Currently, neutrino detectors can provide three important pieces of information about reactors:**

1. **Reactor state (on/off):** Neutrino emissions are much higher when a reactor is operating. A neutrino detector can detect a reactor turning on or off from a distance.

2. **Reactor power:** Measuring the rate of neutrino emissions from a reactor reveals the reactor's power level in real time.

3. **Fissile content of core:** Observing the rate and energy spectrum of neutrino emissions from a reactor over time can provide information about the core contents, such as removal of plutonium from the core.

**With further research and development, neutrino detectors could provide the following information:**

- **Isotope production in reactors:** Neutrino detectors could look for the distinctive signals of isotope production technology, including plutonium breeding blankets and tritium production via lithium bars.

- **Irradiated fuel:** After removal from a reactor, fuel continues to produce low-level neutrino emissions, which could be monitored in fuel storage facilities.

- **Post incident state of a reactor facility:** After an accident, a neutrino detector could provide information about the state of the reactor core and facility.

## Demonstrated neutrino detection systems

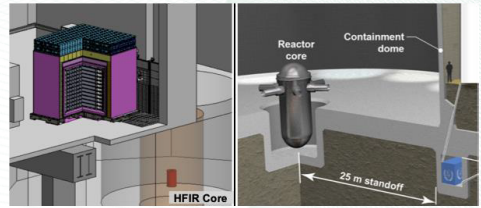

**PROSPECT**
Size: 4 tons
Location: Above ground
Distance: ~8 m
Reactor: Research reactor
(Credit PROSPECT collaboration)

**SONGS**
Size: 0.7 tons
Location: Below ground
Distance: ~25 m
Reactor: Single power reactor
(Credit SONGS Collaboration)

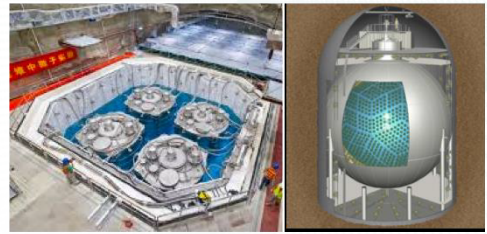

**Daya Bay**
Size: 20 tons
Location: Below ground
Distance: ~1.7 km
Reactor: Multiple power reactors
(Credit Daya Bay Collaboration)

**KamLAND**
Size: 1,000 tons
Location: Below ground
Distance: ~175 km
Reactor: Multiple power reactors
(Credit KamLAND Collaboration)

Note: The PROSPECT system works on the earth's surface, and similar systems could be deployed on a mobile platform. The other three detector technologies require an underground site.

**Compared to other reactor monitoring tools, neutrino detectors have these advantages:**

- Reactor power and fissile content can be monitored *without* operator declarations of reactor power, operating history, or refueling schedule.

- Detectors are always located outside of the reactor building, so no connection to plant facilities is required. Consequently, they are minimally invasive.

- There are no known ways to shield, suppress, or fake a neutrino signal.

- Unattended and remote operation is normal for this technology.

## Contact

**Nathaniel Bowden**
Staff Scientist
Lawrence Livermore National Laboratory
925.422.4923
bowden20@llnl.gov

**Michael Foxe**
Staff Scientist
Pacific Northwest National Laboratory
509-375-2053
Michael.Foxe@pnnl.gov

**Jason Newby**
Staff Scientist
Oak Ridge National Laboratory
865.241.2164
newbyrj@ornl.gov

**Patrick Huber**
Professor
Virginia Tech
540.231.8727
pahuber@vt.edu

**Limitations of neutrino detection technology**

- Very long-range monitoring (hundreds of kilometers) would require very large detectors, so shorter distances are more practical. Shorter distances will require permission from the reactor operator or host country for deployment.

- Neutrino detectors can be rendered inoperable in many ways, most of which are similar to any other active monitoring device (e.g. cameras, deployed on-site at a nuclear facility).

- The cost is relatively high compared to existing methods.

## Cost estimates

- $1–2M per ton for surface detectors

- $5-10M per 10 ton below ground liquid scintillator

- $50-100M per 1000 ton below ground water detector

- Plus deployment specific costs

## Neutrinos

Neutrinos are practically massless, electrically neutral, stable particles. Nuclear reactors and associated materials, like spent nuclear fuel or reprocessing waste, emit electron antineutrinos in beta decays. For brevity the term "neutrino" is used throughout with the understanding that these are electron antineutrinos. The three interaction channels, ordered by their relevance for applications, are: inverse beta decay (IBD), electron scattering (ES) and coherent elastic neutrino nucleus scattering (CEvNS).

## Reactor Neutrino Emissions

Neutrinos originate from the beta decays of neutron-rich fission fragments and on average 6 neutrinos per fission are produced with 2 of them being able to induce IBD. A reactor of 1 GW thermal power produces approximately $10^{20}$ neutrinos per second; a 1 kg detector at a distance L=10m from reactor with thermal power P=1 GW results in 4,000 IBD reactions per year $4000/10 \text{ m}^2 L^2 P \text{GW}$.

The fission fragment distribution depends on which isotope is undergoing fission, therefore the aggregate neutrino emissions also vary in total number and energy spectrum. For example, fission of plutonium-239 results in a softer (lower average energy) neutrino spectrum than fission of uranium-235. This isotopic effect is preserved in all neutrino interaction modes but requires collection of sufficient event statistics to be utilized.

## Types of Neutrino Interactions

There are three primary neutrino interactions. The first has been extensively demonstrated by experiments at nuclear reactors while the other two have limited or no demonstrations at reactors.

## 1. Inverse Beta Decay (IBD)

In IBD a neutrino interacts with a free proton (hydrogen nucleus) and produces a positron and a neutron, where the positron carries almost all of the kinetic energy of the neutrino and the neutron carries almost all of the momentum. Neutrons are heavier than protons and thus there is a minimum reaction (threshold) energy of 1.8MeV required. The positron results in a prompt energy deposition, whereas the neutron will be captured, once it thermalizes after 10-200 microseconds, allowing for a delayed coincidence detection. IBD detectors are based on organic scintillators, but water has also been proposed. To date only IBD has yielded signals with characteristics suitable for applications. The IBD cross section weighted over the reactor neutrino spectrum is approximately $6 \times 10^{-19}$ barn; for reference, 1 barn is a typical neutron scattering cross section on hydrogen.

## 2. Elastic Electron Scattering (ES)

In ES a neutrino (of any type) scatters off and imparts recoil energy to an atomic electron. This reaction can happen at any neutrino energy, i.e. it is threshold-less, but the recoil energy decreases with neutrino energy. For a water detection medium, the effective cross section for ES averaged over the reactor neutrino spectrum is $1.7 \times 10^{-19}$ barn. For neutrinos of energies significantly larger than the electron mass of 511 keV the recoil electron approximately preserves the neutrino direction. This reaction has no other signatures that can suppress background, but lends itself well to the use in large-scale water Cerenkov detectors. ES with directional information has been observed for neutrinos from the Sun down to 3.5 MeV but not yet with reactor neutrinos.

## 3. Coherent Elastic Neutrino Nucleus Scattering (CEvNS)

In CEvNS a neutrino (of any type) scatters off a nucleus and transfers recoil energy to it. This reaction can happen at any neutrino energy, i.e. it is threshold-less, but the recoil energy decreases with neutrino energy and is typically very low compared to common background sources. The signature of a recoiling nucleus is a very high specific energy loss, but like ES there is only one detectable particle produced making background suppression difficult. The reaction cross section is proportional to the square of the number of neutrons, which for heavy nuclei leads to a significant enhancement and cross sections as large as $10^{-15}$ barn per target nucleus. However, per unit detector mass the gain relative to IBD is at most a factor of 100. This reaction has been observed for the first time in 2017 with neutrinos of 50MeV energy, and it has not yet been observed for the more challenging case of reactor neutrinos which characteristically have less than 10MeV energy.

# Monitoring Reactor Power with Neutrino Detectors

Monitoring a reactor's power output is essential for operational control and can provide information about material fission history during a crucial stage of the nuclear fuel cycle. Information about a material's fission history is useful for *nuclear materials accounting*, a nuclear safeguards inventory process that ensures all special nuclear material at a site is controlled and accounted for.

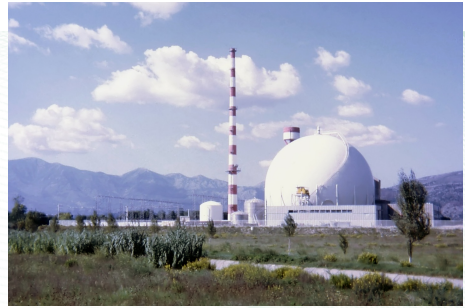

Existing nuclear safeguards programs for nearly all reactors do not exploit reactor power information; instead, fissile material production is monitored using procedural controls, containment and surveillance, and indirect measurements of spent fuel. Although commercial and research reactor operators collect thermal power information through thermohydraulic measurements, these methods may not be applicable for emerging reactor designs. Monitoring a reactor's power through its neutrino emissions is a noninvasive approach that can benefit both reactor operations and nonproliferation efforts.

## Safeguards applications

Detection rates in a neutrino-based reactor power monitor are roughly proportional to the reactor's thermal power divided by the detector's standoff distance squared. Neutrino detection technology supports the following safeguards activities:

a. Determining the presence or absence of a reactor

b. Detecting a change in the reactor state (on/off)

c. Recording reactor power with some accuracy over time

All three activities support nuclear safeguards, but only the third application, recording the power over time, provides operational context. The number of detected events needed increases from simply determining a reactor is present (a) to recording its activities over time (c). Measurements at a distance greater than 100 km are only useful for determining if the reactor is present (a). Inferring a reactor's operational status, (b) and (c), must be done at significantly smaller distances, so these require cooperation with the reactor operator. The typical detector size for learning about the reactor's operational status, (b) and (c), is 1–100 tons.

All three safeguards applications for neutrino detectors have been experimentally demonstrated by basic and applied science experiments. Percent-level accuracy for daily reactor power can be

and has been recorded over years-long time scales, with a sensitivity independent of reactor type and improving with increasing reactor power. With a 4 ton detector[1] at 20 m, the on/off transition of a 100 MW reactor can be observed within 1 day.[2] Detection time increases to 2 weeks for a 20 MW reactor. Such a detector could be deployed over a days-long time scale inside a standard shipping container with minimal infrastructure or at an indoor storage location well-removed from primary reactor operations. For a 30 ton detector at a 1 km distance deployed 100 m underground, the change of on/off state in a 20 MW thermal reactor can be detected within 250 days. This time decreases rapidly with increasing reactor power. For a 100 MW reactor power, the time shrinks to 15 days. The second scenario requires site excavation and detector assembly on-site.

[1] Realistic background measurements based on existing experiments and appropriate for the specific detector and overburden (for shielding purpose) is used in all cases. All detector masses given are based on demonstrated detection efficiencies.

[2] Quoted sensitivities here based on a false positive rate of 1%–5% depending on the specific case and a fixed true positive rate of 95%.

*Chairs*

**Bryce Littlejohn**
Assistant Professor
Illinois Institute of Technology
blittlej@iit.edu

**Anna Erickson**
Associate Professor
Georgia Institute of Technology
Anna.erickson@me.gatech.edu

**Nathaniel Bowden**
Staff Scientist
Lawrence Livermore National Laboratory
bowden20@llnl.gov

# Fissile Content of Nuclear Reactors

Detecting the diversion or undeclared production of nuclear materials is a primary goal of nuclear safeguards, so knowing the fissile content of a reactor core is important information for safeguards efforts.

Current reactor safeguards implementations use a combination of nuclear material accountancy, nondestructive and destructive measurements of fuel, and containment and surveillance. The combination of pre- and post-irradiation measurement of reactor fuel and predictions of fuel activation and depletion from modelling, using the declared operating history, can be used to validate that history after the fact and produce an indirect estimate of fuel fissile content.

Future reactor types that use fuel dissolved in the coolant, such as liquid metal and molten salt reactors, pose a considerable safeguards challenge. Conventional accountancy and containment and surveillance techniques, based on tagging discrete fuel elements, will not be possible. Fissile materials accounting, based on chemical analysis of molten salts or liquid metals, introduces a new proliferation pathway through sample collection.

Neutrino measurements can provide a continuous measurement of reactor fissile content. The energy spectrum of neutrinos is sensitive to the specific mix of fissionable isotopes in the reactor. These characteristic energy changes have been theoretically predicted[1] but only recently experimentally observed.[2] Inference of reactor fissile content using antineutrino measurements requires a relatively high counting rate to achieve the necessary statistical uncertainty. Therefore, a neutrino detector would have to be close to the reactor, likely less than 1 km, which would require cooperation of the reactor operator. Any configuration that can measure the core fissile content will also provide an accurate measurement of reactor power.

In terms of absolute plutonium mass, the sensitivity is best for reactor types with a high fission density, such as traditional pressurized light water moderated designs, and decreases for decreasing fission density, such as natural uranium–fueled graphite-moderated designs. To determine absolute plutonium mass, this measurement becomes more difficult as the reactor thermal power increases because the plutonium content increases with power.

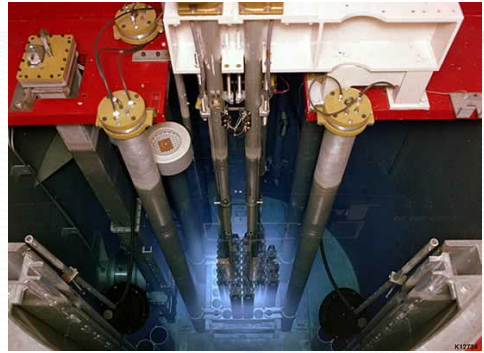

Diversion of 8 kg of plutonium in a 100 MW light water reactor can be detected by a 20 ton detector at a distance of 20 m within 200 days without information about the reactor's operating or refueling history.[3] In this example, the detector system would fit inside a standard shipping container and could be deployed with minimal infrastructure aboveground within a very short period of time (days), assuming that the detector system has been assembled off-site.

*Chairs*

**Jon Link**
Professor of Physics
Virginia Tech
jmlink@vt.edu

**Nathaniel Bowden**
Staff Scientist
Lawrence Livermore National Laboratory
bowden20@llnl.gov

nutools.ornl.gov

# Non-Fission Material Transmutation

## Neutrino detection

For each beta decay, a corresponding *neutrino* is also emitted. This panel will explore the detection of neutrinos from reactor materials other than fuel, such as the production of weapons, medical, or industrial isotopes.

Detecting neutrinos from a nuclear reactor largely involves measuring the decay of fission products. Neutrinos can also be generated from nuclear reactions besides fission, which may happen both intentionally or inadvertently. These reactions include, but are not limited to, reactor production of plutonium via breeding blankets, tritium via lithium bars, or various industrial or medical isotopes. These production mechanisms create lower energy neutrinos at significantly reduced numbers compared to the fission of a power reactor. Consequently, detecting these production activities requires detection technologies that have not yet been implemented at nuclear reactors.

## Non-Fission Transmutation

*Non-fission material transmutation* broadly refers to the elemental or isotopic change of material in a reactor, either intentionally or as byproducts, through nuclear processes other than fission. Significant material transmutation can occur within a variety of design components in a reactor. Generally, the reactor monitoring applications commonly discussed rely on the proportionality of detected neutrino rate to fissions (i.e., power level).

Although fission reactions ultimately yield the bulk of the neutrinos, other neutron-induced interactions associated with transmutation can, under certain situations, produce a significant, and potentially detectable, number of non-fission-derived neutrinos. The contribution of these reactions to heat production is small compared to fission (i.e., contribution to power level), but they can become important as the required precision of neutrino production predictions and subsequent reactor monitoring is increased.

## Safeguards considerations

The most prominent transmutation is the production of plutonium using a breeding blanket,[1,2] which requires additional safeguards considerations. Transmutations produce fissile plutonium isotopes without contributing significantly to the power level. Reactors

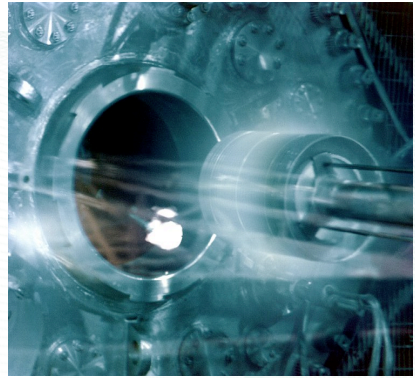

configured in this way are called *breeders* because they produce more fissile content than they consume. The extent to which these transmutations can be detected via their associated neutrinos needs to be explored further, though studies suggest that recently demonstrated coherent scattering neutrino detection technologies may provide a viable pathway to realizing this capability.[3,4] A similar situation includes the detection of nuclear reactor production of tritium via lithium transmutation.[5,6]

Monitoring reactors for production of other isotopes of interest, such as medical or industrial isotopes, has yet to be explored. In addition to the technical challenges of lower signal rate and energy threshold compared to fission products, variations in reactor designs have the potential to complicate predictions of non-fission material transmutations and thus detection confidence. Nonetheless, the detection of non-fission-related neutrinos for applications other than power monitoring has yet to be deeply explored, and defining the potential advantages of the capability for traditional safeguards measures in this area is necessary.

*Chairs*

**Andrew Conant**
Postdoctoral Research Associate in Nuclear Engineering
Oak Ridge National Laboratory
conantaj@ornl.gov

**Pieter Mumm**
Physicist
National Institute of Standards and Technology
hans.mumm@nist.gov

**Jason Newby**
Staff Scientist
Oak Ridge National Laboratory
newbyrj@ornl.gov

# Regional Reactor Discovery, Exclusion, and Monitoring

This panel will explore the prospects for neutrino detection to benefit two remote proliferation detection use cases: discovery of undeclared, research-scale nuclear reactors and verification of the operation and monitoring of known nuclear reactors. These capabilities are sought for reactor–detector distances that exceed 2 km.

Small (tens of megawatts) undeclared nuclear reactors can produce plutonium at a high enough rate to support clandestine nuclear weapons programs. Consequently, their discovery and exclusion in a regional context is a high priority for nuclear nonproliferation. Unverified operation of declared nuclear reactors presents similar nonproliferation concerns. Neutrino-based methods may expand the existing technical tool set for reactor discovery, exclusion, and monitoring by exploiting a characteristic signature of fission that is immune to shielding and spoofing.

Compared to existing methods for remote reactor observation, neutrino detectors offer unique features that may be of use in current or future monitoring activities. The existing tools and technologies exhibit limitations such as intermittent operation, unpredictability in the efficacy of data collection and source term magnitude, limited geographical coverage, or inability to provide tight constraints on the reactor location. By contrast, unique features of neutrino detectors include: persistence; the ability to detect or exclude reactor activity in a wide geographical region without external cueing information; insensitivity to weather, shielding and other environmental factors; the potential to place constraints on, or directly measure, the operational status and total thermal power of the reactor and thereby estimate the maximum possible rate of plutonium production in the discovered reactor.

The technology has already been demonstrated over the 2–20 km range in existing underground scientific experiments and could be adapted for monitoring and exclusion applications with little or no design modifications required. Challenges for long-range reactor discovery, exclusion, and monitoring using neutrino detectors include the intrinsically low signal rate and the need to suppress both the neutrino and non-neutrino backgrounds. Because of the low neutrino interaction rate, discovering a 50 MW reactor within a year from 1,000 km distance would require a 335 kt detector,[1] provided that such a detector can reject the neutrino backgrounds from

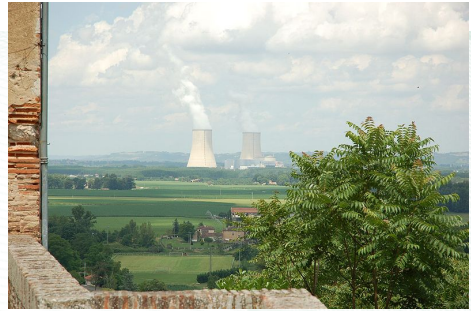

existing reactors. The largest existing neutrino detector, Super Kamiokande, has an active volume of about 25 kt and cost about $100M to build.

*Chairs*

**Igor Jovanovic**
Professor
University of Michigan
ijov@umich.edu

**Michael Foxe**
Staff Scientist
Pacific Northwest National Laboratory
Michael.Foxe@pnnl.gov

nutools.ornl.gov

# Post-Accident Reactor Monitoring

This panel will explore the application of neutrino detectors, which have been demonstrated as reactor monitors at full power, to scenarios involving fuel signatures that could signal a reactor accident or a transient event.

Neutrino detectors have demonstrated the capability to monitor reactor operation, including status, power level, and fissile inventory. All demonstrations have focused on reactors operating in steady state at full power. The technology has improved so that accident scenarios, in which there may or may not be sustained fission source of neutrinos, could now be considered. Challenges and considerations for this application include signal rates, background rejection, potential physical translation of fuel, and resilience to adverse or severe conditions.

Nuclear reactors are designed to operate under normal conditions as well as under certain accident conditions, which occur with an anticipated frequency and can include fuel damage, containment integrity, or radiation release off-site. Real-time information about a nuclear reactor after an accident can be crucial to maintaining the integrity of the reactor and radiological safety of the area. If an accident is known to have occurred, more information about the extent of the contamination is needed. Because neutrino detectors detect the by-products from fission, a sustained signal could be indicative of a continuing chain reaction that has yet to be brought under control.

The International Atomic Energy Agency would like to monitor reactivity for an extended period of time after an accident, and current neutron monitors may experience harsher than normal operating conditions or calibration challenges.[1] Having a real-time method of assessing whether fuel changed state is desirable if an accident remains in a critical configuration.

In the case of the Fukushima accident, the fuel melted, and traditional instrumentation was not available or useful.[2] A major challenge of neutrino detection in this application is the small magnitude of the signal compared to full power operation. The detection has flexibility in operating modalities (e.g., permanently emplaced or mobile), although the latter will have background rejection challenges depending on the distance. The extent to which neutrino detectors are applicable under a wide range of accident scenarios needs to be investigated (e.g., the levels of radioactivity may be so high that operation of neutrino detectors is difficult).

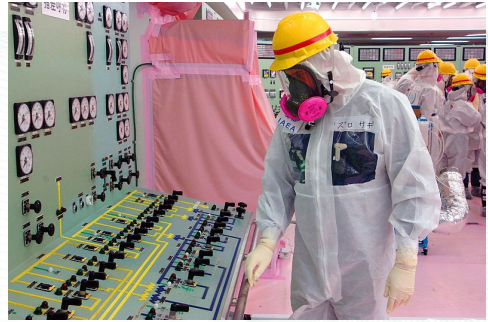

[1] "Accident Monitoring Systems for Nuclear Power Plants." International Atomic Energy Agency. 2015.
[2] M. Fackler. "Six Years After Fukushima, Robots Finally Find Reactors' Melted Uranium Fuel." *New York Times*. November 19, 2017.

*Chairs*

**Andrew Conant**
Postdoctoral Research Associate in Nuclear Engineering
Oak Ridge National Laboratory
conantaj@ornl.gov

**Jason Newby**
Staff Scientist
Oak Ridge National Laboratory
newbyrj@ornl.gov

nutools.ornl.gov

# Spent Fuel Monitoring

This panel will explore the prospects of neutrino detection for monitoring spent fuel, which has applications in verification of isotopic composition, reprocessing efforts, and nuclear archaeology.

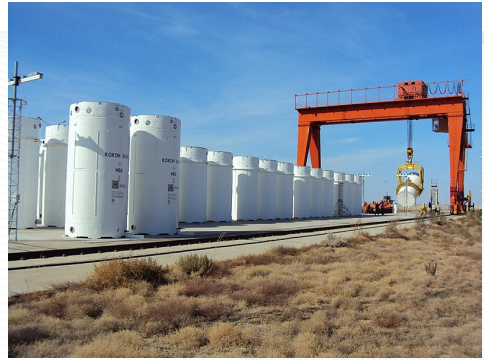

Fewer neutrinos are emitted from spent fuel than from operating reactors. The time scale and various applications have been studied from days after irradiation to long-term storage in a geological repository, but only the former has been measured to date. The low signal-to-background ratio poses a significant challenge in the development of this technology, and future research and development is necessary to further this application.

Neutrino emissions from fuel post-irradiation declines very quickly, within minutes, to a small fraction of the rate during irradiation with the highest energy neutrinos vanishing the fastest. Twenty-four hours after irradiation, only a handful of fission fragment isotopes emit neutrinos above inverse beta decay threshold, which constitutes the limit of our ability to detect neutrinos at reactors. On longer time-scales, only strontium-90, which has a half-life of 29 years, remains with neutrino emissions above the inverse beta decay threshold. Strontium-90's decay chain can produce neutrinos up to 2.2 MeV energy. The fission yield of strontium-90 is around 5%, so it is copiously produced and notably retained in the aqueous phase of the PUREX process. Therefore, also reprocessing waste will exhibit significant neutrino emission because about 1 mol (90 g) of strontium-90 ends up in the waste stream for about 4 kg of separated plutonium. This amount of strontium-90 would result in about 25 events per year in an ideal 5 ton inverse beta decay detector at a of 10 m. The half-life of strontium-90 is long enough that even the oldest spent fuel, dating to 1943, still contains 16% of its original strontium content.

## Scenarios

1. Long-term monitoring of geological spent nuclear fuel repositories, such as at the Yucca Mountain Nuclear Waste Repository.[1]

2. Verification of dry-storage casks.[2]

3. Locating reprocessing wastes in cleanup efforts at known plutonium production sites, like the Hanford Site.[3]

4. Nuclear archeology—After denuclearization, a complete understanding of all past plutonium production is desirable

and neutrino emission from buried reprocessing waste can, in principle, provide an estimate of total plutonium production at a given site. [4]

The challenge in all cases is that event rates are relatively low compared to a running reactor, and the neutrino energy is quite low, accentuating the issue of random backgrounds from natural radioactivity. To date the only actual detection of post-irradiation neutrinos has taken place on a time-scale of days after irradiation. Scenario 1 can be addressed with current detector technology, using single-volume large scale (thousands of tons) liquid scintillator detectors buried deep underground. For scenario 4 scaling, from demonstrated detector performance at the surface without overburden, indicates that reprocessing waste corresponding to 80 kg of separated plutonium could be detected in less than 2 years with a detector which fits inside a standard shipping container. Scenarios 2 and 3 seem to be more challenging and may require further detector research and development. In particular, directional neutrino detection in ton-scale detectors would greatly enhance capabilities for those two cases.

*Chairs*

**Patrick Huber**
Professor
Virginia Tech
pahuber@vt.edu

**Michael Foxe**
Staff Scientist
Pacific Northwest National Laboratory
michael.foxe@pnnl.gov

nutools.ornl.gov

# Neutrino Detection Scientific Engagement

Beyond a technical role as reactor monitors, neutrino detectors offer opportunities to build trust with adversaries, reemploy former weapons scientists, and connect the intellectual resources of the basic science community with nuclear security challenges.

These opportunities arise from an application in which neutrino detectors have already proven useful during the past 60 years: as collaborative tools for science. From a small experiment run by US weapons lab scientists, neutrino physics has grown to a multibillion-dollar venture linking thousands of physicists in the United States, Europe, Russia, China, South Korea, and elsewhere.

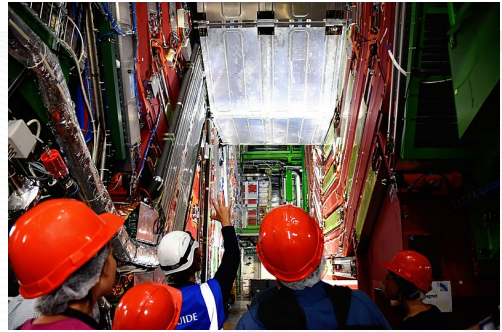

## Opportunities for engagement

Connections to cutting-edge science and to a global community of physicists are special assets that neutrino detectors bring to the Department of Energy's Office of Defense Nuclear Nonproliferation mission. These assets offer utility to the Office of Defense Nuclear Nonproliferation and other nonproliferation agencies in multiple ways:

- Cooperatively fielding a neutrino detector, especially at a former military reactor, could be a low-stakes way to help build trust between the United States and another nation. US agencies have relied on technical projects to help build trust with former adversaries since the Cooperative Threat Reduction program in the former Soviet Union. More recently, neutrino projects have been suggested as one part of "a broader opening of scientific engagements" with Iran.[1] Other cooperative opportunities for neutrino detectors could also arise in the future.[2] In general, neutrino detectors are well-suited to cooperative exercises because they are a novel, militarily-insensitive, and somewhat remotely deployable tool. Because the neutrino physics community spans many nations such a project could be supported multilaterally.

- Cooperative neutrino projects could help connect an adversary's former weapons scientists to nonmilitary work. Directing former weapons scientists to peaceful occupations, rather than work in another weapons program, was one aim of the original Cooperative Threat Reduction program. Officials have also emphasized this objective for North Korea and Iran.

- A cooperative, neutrino-based reactor monitoring project would be a gateway for technical personnel to enter the international particle physics community.

- Applied neutrino projects could help connect scientists and students from the particle physics community with challenges in the US nuclear security enterprise. In particular, these projects can help attract graduate students to security careers.

[1] Joint Comprehensive Plan of Action Annex 3 - Civil Nuclear Cooperation.
[2] R. Carr et al., Science & Global Security 27 (2019).

*Chairs*

**Rachel Carr**
Stanton Nuclear Security Fellow
Massachusetts Institute of
Technology
recarr@mit.edu

**Nathaniel Bowden**
Staff Scientist
Lawrence Livermore National
Laboratory
bowden20@llnl.gov

nutools.ornl.gov



\end{document}

%% file: execSummary.tex
\label{execSummary}

For decades, physicists have used neutrinos from nuclear reactors to advance basic science. These pursuits have inspired many ideas for application of neutrino detectors in nuclear energy and security. While developments in neutrino detectors are now making some of these ideas technically feasible, their value in the context of real needs and constraints has been unclear. This report seeks to help focus the picture of where neutrino technology may find practical roles in nuclear energy and security. 

This report is the final product of the Nu Tools study, commissioned in 2019 by the DOE National Nuclear Security Administration (NNSA) Office of Defense Nuclear Nonproliferation Research and Development (DNN R\&D). The study was conducted over two years by a group of neutrino physicists and nuclear engineers. A central theme of the study and this report is that useful application of neutrinos will depend not only on advancing physics and technology but also on understanding the needs and constraints of potential end-users.

The \hyperref[studyIntro]{\emph{Study Approach}} emphasized broad end-user engagement. The major effort, undertaken from May to December 2020, was a series of engagements with the wider nuclear energy and security communities. Interviews with 41 experts revealed points of common understanding, which this report captures in three \ChapRefCCFindings{}, a \ChapRefFramework{}, and seven \ChapRefUCFindings{}.
The report concludes with two \hyperref[recommendations]{\emph{Recommendations}}. The findings and recommendations are summarized below. The respective ordering within each category does not represent a prioritization or implied value judgement.

\section*{Cross Cutting Findings}

Three findings of this study apply across all potential applications of neutrino technology: \\

\Copy{CCF1}{\noindent \textbf{\hyperref[finding:EndUser]{\emph{End-User Engagement:}}}  The neutrino technology R\&D community is only beginning to engage attentively with end-users, and further coordinated exchange is necessary to explore and develop potential use cases.}  \\

\Copy{CCF2}{\noindent \textbf{\hyperref[finding:TechnicalReadiness]{\emph{Technical Readiness:}}} The incorporation of new technologies into the nuclear energy or security toolbox is a methodical process, requiring a novel system such as a neutrino detector to demonstrate sufficient technical readiness.} \\

\Copy{CCF3}{\noindent \textbf{\hyperref[finding:Siting]{\emph{Neutrino System Siting:}}} Siting of a neutrino-based system requires a balance between intrusiveness concerns and technical considerations, where the latter favor a siting as close as possible.}

\section*{Use Case Findings}

Seven findings of this study pertain to specific use cases discussed during conversations with the wider nuclear security and nuclear energy communities. This report evaluates these hypothetical use cases using a common framework consisting of four criteria: the need for a new or improved capability in a particular application space; the existence of a neutrino signal; the availability of a neutrino detection technology; and the compatibility of that technology with end-user implementation constraints including cost, workforce requirements, timelines, and other logistical considerations. The \ChapRefUCFindings{}, presented with full analysis later in the report, are briefly summarized as:  \\

\Copy{UCF2}{\noindent \textbf{\hyperref[finding:CurrentIAEA]{\emph{Current \IAEA\ Safeguards:}}}  For the vast majority of reactors under current IAEA safeguards, the safeguards community is satisfied with the existing toolset and does not see a specific role for neutrinos.} \\

\Copy{UCF4}{\noindent \textbf{\hyperref[finding:AdvancedReactors]{\emph{Advanced Reactors:}}} Advanced reactors present novel safeguards challenges which represent possible use cases for neutrino monitoring.} \\

\Copy{UCF1}{\noindent \textbf{\hyperref[finding:FutureDeals]{\emph{Future Nuclear Deals:}}}  There is interest in the policy community in neutrino detection as a possible element of future nuclear deals involving cooperative reactor monitoring or verifying the absence of reactor operations.} \\

\Copy{UCF3}{\noindent \textbf{\hyperref[finding:ReactorOperations]{\emph{Reactor Operations:}}} Utility of neutrino detectors as a component of instrumentation and control systems at existing reactors would be limited.} \\

\Copy{UCF5}{\noindent \textbf{\hyperref[finding:NonCoop]{\emph{Non-Cooperative Reactor Monitoring or Discovery:}}} Implementation constraints related to required detector size, dwell time, distance, and backgrounds preclude consideration of neutrino detectors for non-cooperative reactor monitoring or discovery.}\\

\Copy{OAF8} {\noindent \textbf{\hyperref[finding:SpentFuel]{\emph{Spent Nuclear Fuel:}}} Non-destructive assay of dry casks is a capability need which could potentially be met by neutrino technology, whereas long-term geological repositories are unlikely to present a use case.} \\

\Copy{OAF7}{\noindent \textbf{\hyperref[finding:PostAccident]{\emph{Post-Accident Response:}}} Determining the status of core assemblies and spent fuel is a capability need for post-accident response, but the applicability of neutrino detectors to these applications requires further study.}

\section*{Recommendations}

In light of the study findings, this report makes two recommendations to the sponsor which together present a pathway to practical use of neutrino technology in service of policy needs. The \hyperref[recommendations]{\emph{Recommendations}}, each expanded upon later in the report, are: \\

\Copy{Rec1} {\noindent {\bf \hyperref[rec:EngagementSupport]{\emph{Recommendation for End-User Engagement:}}} DNN should support engagement between neutrino technology developers and end-users in areas where potential utility has been identified.} \\

\Copy{Rec2} {\noindent {\bf \hyperref[rec:Portfolio]{\emph{Recommendation for Technology Development:}}} DNN should lead a coordinated effort among agencies to support a portfolio of neutrino detector system development for areas of potential utility, principally in future nuclear deals and advanced reactors. }\\

%% file: studyIntro.tex
\label{studyIntro}

\section{Context of this Report}

This report is the final product of the Nu~Tools study, commissioned in 2019 by the DOE National Nuclear Security Administration (NNSA) Office of \DNNRD. The study was performed by the Nu~Tools Executive Group, a team of neutrino physicists and nuclear engineers from U.S. universities and U.S. government laboratories. \DNNRD\ charged the group “to facilitate broad engagement with interested communities on the topic of antineutrino\footnote{This document uses the more general term "neutrino" to refer to both neutrinos and antineutrinos, except when quoting a source that uses the specific term "antineutrino". The difference in terminology is not significant for the present discussion.}-based monitoring of nuclear reactors and associated post-irradiation fuel cycle activities,” with a focus on “\textbf{potential utility} of antineutrino detection technologies and required detection capabilities... in the context of existing or potential policy needs” (emphasis from the original charge\footnote{The Charge to the Executive Group of the Antineutrino Reactor Monitoring Scoping Study is provided in \autoref{charge}.})

In response to the sponsor's request, this report seeks to provide strategic input to guide possible future R\&D investments in the \DNNRD~
portfolio. This public report also seeks to inform the R\&D efforts of scientists and engineers interested in neutrino applications. Finally, the report offers members of the nuclear energy and nuclear security communities a perspective on where neutrino technology could eventually have practical value for them. This document does not attempt to provide a comprehensive technical primer or survey of relevant literature. For these, the reader is referred to a recent survey of applications-oriented neutrino technology\autocite{RMP}. Appendix \ref{glossary} provides a glossary of technical terms used in the present report.

\section{Focusing on Utility }

The Nu Tools Executive Group was tasked with evaluating the utility of neutrino detectors in the context of existing or potential nuclear energy or nuclear security needs.
The Executive Group recognizes that utility, defined as detector deployment to meet a nuclear energy or nuclear security need, is only one of the many valuable outcomes that a neutrino R\&D program might produce. Broader benefits include nuclear workforce development, international cooperation among nuclear agencies, and development of new scientific knowledge and technical advances. 
A foundation in utility allows R\&D programs to better capitalize on these broader impacts.
That is, a neutrino detector pursued for nuclear energy or security applications must be realistically deployable to serve as an effective platform for workforce training and international cooperation. Although the focus of this report is utility, broader benefits are noted throughout the report where possible.

With its focus on utility, this study adds practical context to the prior literature on neutrino applications. Most earlier studies have focused on characterizing neutrino signals and possible detection technologies. This study recognizes that utility also depends on the needs and constraints of end-users: reactor designers, inspectors, diplomats, and other specialists.

\section{Engagement with Expert Communities}
\label{sec:Engagement}
\begin{figure}
    \centering
    \includegraphics{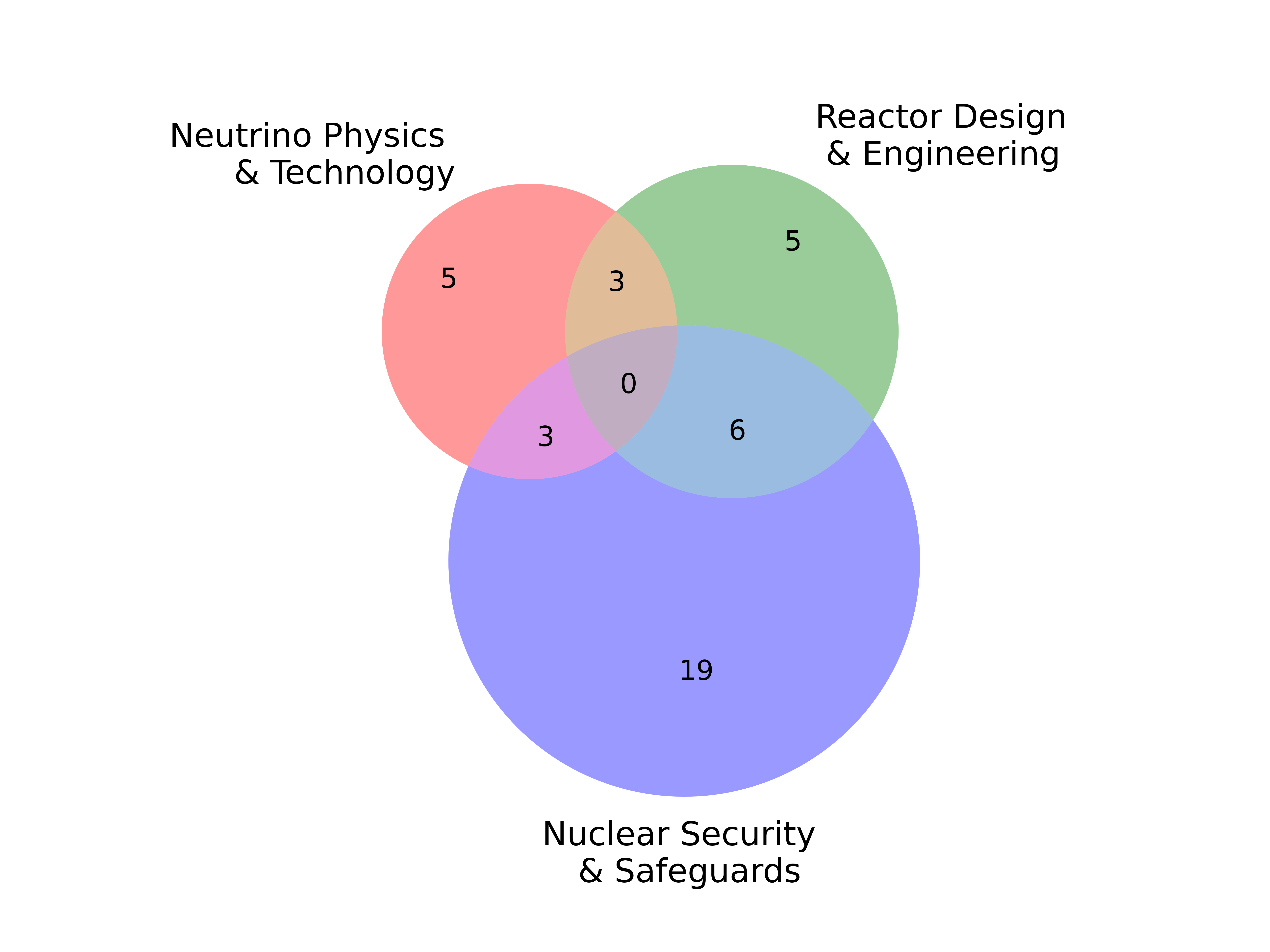}
    \caption{Diagram depicting the fields of specialization of the 41 experts engaged for this study. Interviewees were assigned to each set based on their area of expertise as identified by the Executive Group. Not represented in the diagram are an additional set of engagements with the neutrino technology development community, conducted via a mini-workshop in July 2020.}

    \label{fig:venn}
\end{figure}

To collect expert views on practical considerations, the Nu~Tools study prioritized broad engagement with relevant communities, as shown in Figure~\ref{fig:venn}.
Community assessment conducted by members of the Nu~Tools Executive Group was performed through semi-structured interviews and a mini-workshop. Interviewees were selected by Executive Group members with an emphasis on experts outside the physics research community, including international and domestic safeguards practitioners, nuclear reactor vendors and operators, and nuclear policy experts with experience in government agencies and non-governmental organizations. Most interviewees were known to Executive Group members through previous contacts.
A list of all interviewees is shown in \autoref{ch:interviewee_list}.

To survey the scientific community, a mini-workshop was held in July 2020. Specialists in neutrino technology development, including participants in the Applied Antineutrino Physics (AAP) conference series, were invited to provide their assessment of the utility of neutrino technologies for nuclear energy and security.
Details about the mini-workshop appear in \autoref{miniWorkshop}.  

As background material for the interviews and mini-workshop, the Executive Group developed fact sheets 
covering a broad overview of potential neutrino applications. The fact sheets appear in \autoref{FactSheets}. These sheets include generalizations of use cases that have been examined in the literature or discussed in the technology development community. Overall, the fact sheets offered an inclusive starting point with context relevant to a variety of potential end-users. The material was largely informed by known physics and conceivable detector technology. The fact sheets covered the following areas:

\begin{itemize}
    \item \textit{Reactor power monitoring}: determining the presence or absence of a reactor, a change in the reactor state, or tracking the reactor power over time.
    \item \textit{Fissile content tracking}: continuously measuring the reactor fissile content, particularly its plutonium inventory.
    \item \textit{Non-fissile material transmutation}: production of plutonium, tritium, or various industrial or medical isotopes through nuclear processes other than fission. 
    \item \textit{Irradiated fuel monitoring}: verification of dry-storage casks, long-term monitoring of geological spent fuel repositories, locating reprocessing waste in cleanup efforts, and nuclear archaeology.
    \item \textit{Post-incident monitoring}: detection of criticality in accidents involving fuel damage, containment integrity, or radiation release.
    \item \textit{Regional reactor observation}: remote discovery of undeclared nuclear reactors or verification of the operation and monitoring of known reactors.
    \item \textit{Scientific engagement}: cooperative efforts to build trust with adversaries, reemploy former weapons scientists, and leverage the nuclear security infrastructure to address basic scientific questions.
\end{itemize}

The fact sheets were included with interview invitations, and provided guidance on potential topic areas for discussion as categorized by the Executive Group. However, they served only as conversation starters, and their categorization of use cases does not correspond directly to findings in this report.
Each interviewee also received a set of questions to frame the discussion.
The interview questions were designed to illuminate needs and desires of the end-user community and potential implementation constraints.
At least two Executive Group members participated in each interview.
Accounts of the interviews were documented simultaneously by Executive Group members and later combined into a single account provided to the interviewee for review of accuracy and completeness.

\section{Report Preparation}

Following the May--September 2020 interviews, the Executive Group summarized the major findings, taking special care to synthesize viewpoints and capture community consensus.
The result of this stage of analysis was an interim report provided to \DNNRD. To fill gaps in some topics and clarify unsettled points, more interviewees were identified and interviewed from September--December 2020.
This final report builds upon conclusions from the interim report, a review of all interviews, and systematic analysis by the Executive Group.

%% file: crossCuttingFindings.tex
\label{crossCuttingFindings}

Engagement between the Nu Tools Executive Group and members of expert communities led to the following three cross-cutting findings. These findings apply to all potential use cases for neutrino technology: \\

\section{End-User Engagement}
\label{finding:EndUser}

\input{crossCutting/userPerception}

\section{Technical Readiness}
\label{finding:TechnicalReadiness}
\input{crossCutting/techReadiness}

\section{Neutrino System Siting}
\label{finding:Siting}
\input{crossCutting/siting}

%% file: crossCutting/userPerception.tex
\noindent \textbf{The neutrino technology R\&D community is only beginning to engage attentively with end-users, and further coordinated exchange is necessary to explore and develop potential use cases.}  \\

Although many experts see some potential in neutrino technologies, they do not see a use case that is compelling enough to justify the adoption of neutrino-based technology at this time. For the most part, potential use cases have been identified, developed, and discussed {\it within} the neutrino physics community. In these studies, the primary focus has been on understanding neutrino signals and developing detection technologies.

While these efforts have effectively demonstrated the general features of the technology, this approach has not been successful in establishing enduring connections and credibility with end-users. This has created an impression that neutrino technology developers are advocating for a specific approach without developing a deep understanding of significant real-world goals and constraints, both political and technical, which has limited the exploration of potential use cases with end-users.
Generally, experts agree that neutrino technology advocates have not engaged end-users sufficiently, nor developed the necessary comprehension of their needs and constraints, to develop a mutual understanding between these groups. 

Systematic and sustained two-way exchange between interested neutrino physicists and end-user communities is necessary to identify use cases that meet all four criteria presented in the \ChapRefFramework{}.
Interviewees feel that some of this engagement can be effectively conducted in end-user forums such as the Institute of Nuclear Materials Management Annual Meeting. To strengthen credibility with potential end-users, neutrino experts should use these forums to develop a more sophisticated understanding of end-user needs and constraints and clearly communicate current neutrino detector capabilities. These engagements may be enhanced through real-time, two-way discussions of use cases, such as panel discussions at conferences and in dedicated working groups.\\

%% file: crossCutting/techReadiness.tex
\noindent \textbf{The incorporation of new technologies into the nuclear energy or security toolbox is a methodical process, requiring a novel system such as a neutrino detector to demonstrate sufficient technical readiness.} \\

The political significance of safeguards and verification calls for robust technologies with unambiguous outputs. 
High standards of accuracy and reliability also apply to reactor instrumentation.
Before incorporating a new technology in any nuclear energy or security system, end-users will require it to have achieved an appropriate Technical Readiness Level (TRL).\footnote{See DOE Technology Readiness Assessment Guide, \url{https://www.directives.doe.gov/directives-documents/400-series/0413.3-EGuide-04-admchg1}} 
Generally, this will be higher than the TRL of current neutrino detectors. 
Considering the implementation of neutrino-based measurements, several scenarios are conceivable, {\it e.g.}: (1) integration of a neutrino system with a new facility in the design and construction phase,
(2) a neutrino system tailored to the constraints presented by an existing facility, and (3) a standardized neutrino system that could be used at various existing facilities ({\it e.g.}, siting a mobile neutrino system external to a reactor containment building). 
The specific scenario and requirements of the end-user will define the TRL requirements for a neutrino system.

End-users may be more willing to consider low-TRL systems in the context of facilities based on new technologies where gaps in safeguards or instrumentation capabilities exist, such as in advanced reactor types still under development.
For end-users in the first scenario above, particularly in the context of reactor instrumentation, a neutrino system could enter the facility planning process at a relatively low TRL\@. TRL~4 is sufficient to consider a system as part of a conceptual design review,\footnote{See DOE Order 413.3B: Program and Project Management for the Acquisition of Capital Assets, April 2016, \url{https://science.osti.gov/opa/Project-Management/Processes-and-Procedures/Department-of-Energy}} and TRL~6 is sufficient to move to preliminary and final design reviews. 
For reactor monitoring applications, early incorporation of a neutrino system into facility planning would also align with the principle of safeguards-by-design\autocite{sbd}. 

However, safeguards end-users ({\it e.g.}, those tasked with verifying reactor operator declarations) would typically expect a technology to have been demonstrated to a TRL of 7 to 8 before incorporation in their planning processes. 
That is, they seek a full-scale system prototype demonstration in an operationally relevant environment, potentially with rigorous qualification tests. In addition to proving the technical qualifications of the system, such a demonstration would give end-users the necessary experience in operating a neutrino system and interpreting the data it provides. A neutrino system demonstration of this type is likely a requirement prior to consideration of the technology for the second and third scenarios above, and would bolster consideration in the first scenario.

As a neutrino technology proceeds through successively higher TRLs, end-user input should be an integral part of the process.
This input is necessary to define system requirements in all scenarios identified above. The end-users---safeguards personnel, reactor designers, and/or reactor operators---will provide necessary inputs for determining required capabilities and implementation constraints. 
Planning and conducting a demonstration should be viewed as an opportunity for end-users and technology developers to collaboratively advance a promising concept towards field readiness.

%% file: crossCutting/siting.tex
\noindent \textbf{Siting of a neutrino-based system requires a balance between intrusiveness concerns and technical considerations, where the latter favor a siting as close as possible.}\\

Interviewees in this study view non-intrusiveness as a key advantage of neutrino-based monitoring approaches. This feature is desirable for all  viable use cases considered in this study, including future nuclear deals, advanced reactor safeguards, spent fuel monitoring, and incident response. The concept of intrusiveness includes several aspects when implementing a monitoring technology: concerns that a technology and the physical access required for its installation and operation could cause technical interference or other disruption at a facility, as well as concerns that the technology may reveal sensitive information beyond the monitoring task.

Neutrino-based monitoring can assuage intrusiveness concerns, since no connection to facility process components is required to access a neutrino signal. This inherent characteristic of neutrinos provides considerable flexibility in system site-selection and largely eliminates technical interference concerns. 
The ability to place a neutrino system outside a facility building or even beyond the facility boundary increases deployment flexibility and reduces the potential to disrupt facility operations or to reveal sensitive information.

Intrusiveness considerations could lead a cooperatively monitored party to prefer sites far from the reactor of interest. However, a very strong impetus to negotiate the closest possible deployment site derives from implementation constraints related to neutrino detector size, cost, and construction timeline. These constraints follow from the fact that the number of neutrinos reaching a detection system falls as the inverse  square of the standoff distance, {\it e.g.} the signal rate at 100~m standoff compared to 25~m is  diminished by $1/16^{th}$.\footnote{Neutrino flavor oscillations also affect signal rates at long baselines ($\gtrsim60$ km) but have a negligible impact at the shorter baselines where this study finds potential utility.} Accepting a more distant deployment site will inevitably result in a combination of increased system size, cost, and deployment time and/or a reduction in obtainable signal rate. 

As with any detection system, the information content available from neutrinos increases with signal rate and decreases with background rate. In the regime of small signal and high background, it is only possible to determine whether a reactor is on or off. With higher signal and/or lower background rates, a determination of the reactor power becomes feasible. Finally, at high signal rates, it becomes possible to exploit the neutrino spectrum to determine the fissile material content of the reactor. 

Furthermore, the background rate scales with the detector size (all else being equal) and is approximately independent of standoff, meaning that available information content inevitably degrades at greater distances even faster than the inverse square of the distance. Absent dramatic advances in detection capabilities, the background can only be significantly reduced by the addition of shielding material or overburden above a monitoring system to attenuate cosmic ray particles. Adding shielding to an aboveground system is only practical at the scale of a few meters of material and substantially increases the system footprint. At some point with increasing standoff and detector size, underground deployment is necessary to achieve sufficient background suppression. In turn, underground deployment requires significant underground excavation and construction. Thus, the transition from surface to underground deployment results in a pronounced increase in system cost, a lack of mobility, and an extension of the construction timeline.

%% file: utilityFramework.tex
\label{utilityFramework}

\section{Four Criteria}

To assess the potential utility of neutrino detectors in specific applications, the Executive Group developed a four-criterion evaluation framework. This framing reflects common themes heard in expert interviews across a variety perspectives. A promising use case for neutrino technology fulfills \textit{all four} criteria:

\begin{enumerate}
    \TabPositions{8cm}  
    \item Need for a new or improved capability,
    \item Existence of a neutrino signal,
    \item Availability of a neutrino detection technology, and
    \item Compatibility with implementation constraints.  
\end{enumerate}

Most previous studies of neutrino applications have implicitly focused on the second and third criteria; that is, they were conducted from the viewpoint of the technology development community. By adding the first and fourth criteria, the Nu Tools study adds practical context.

\section{Division of Expertise}

Dividing the concept of utility into four parts helps clarify where different types of expertise are relevant.
The four criteria in the Nu Tools utility framework are evaluated by two different communities as follows:

\begin{enumerate}
    \TabPositions{8cm}  
    \item Need for a new or improved capability \tab $\rightarrow$ \textit{Determined by end-user communities.}
    \item Existence of a neutrino signal \tab $\rightarrow$ \textit{Determined by technology development community.}
    \item Availability of a neutrino detection technology
    \tab $\rightarrow$ \textit{Determined by technology development community.}
    \item Compatibility with implementation constraints \tab $\rightarrow$ \textit{Determined by end-user communities.}
\end{enumerate}

Neutrino physicists, including most members of the Nu Tools Executive Group, are experts on neutrino signals and detectors (criteria 2 and 3) but not on the needs and constraints of the nuclear energy and security enterprises (criteria 1 and 4). Expertise on needs and constraints resides in the nuclear security and nuclear engineering communities. Accordingly, the experts interviewed in the Nu Tools study provide this report's perspective on criteria 1 and 4. 

\section{Evaluating Each Criterion}

When assessing a potential application for neutrino technology, the Executive Group considered each criterion in the utility framework:

\begin{enumerate}

\item \textit{Capability need} is expressed by the user community as a desire for specific detection and/or monitoring capabilities, which either are entirely missing or not as effective as sought. If applicable, the comparison with existing technologies factors into the assessment of this criterion. Different stakeholders may have different needs in the same use case. Considering the need for a capability is often tied to consideration of cost/effort associated with it: for some capabilities, there is a cost beyond which they are no longer perceived as needed; conversely, some capabilities have no associated value even at very small cost. Thus, there is a coupling between this criterion and implementation constraints (criterion 4).

\item \textit{Existence of a neutrino signal} is evaluated by the neutrino technology community. The assessment is based on well-known physics together with an understanding of the specific use case, the latter often being only approximate. For instance, reactors produce large numbers of neutrinos when operating, and there are suitable detection reactions. In contrast, uranium enrichment does not produce any neutrino signatures. 

\item \textit{Availability of a detector technology} is determined  by the neutrino technology community. In a hypothetical use case, the question is whether it is possible to build a detector sensitive enough to  detect the neutrino signature and whether backgrounds can be sufficiently suppressed. In assessing the availability of detector technologies, a wide range of technological maturity is considered adequate to meet this criterion, even if significant R\&D is still required to obtain a system demonstrating a detection. The criterion is not considered satisfied if major, unforeseeable breakthroughs in technology or new discoveries in neutrino physics are required. 

\item \textit{Implementation constraints} are expressed by the user community. They include cost, workforce requirements (both in terms of number of personnel and training), timeliness of the measurement, lead time to deployment, and general logistical constraints. They can also include issues of intrusiveness and satisfactory compatibility with an agreement.  Consideration of this criterion includes a weighing of the urgency of a capability need versus these constraints.

\end{enumerate}

As noted above, a potential neutrino application is considered promising only if \textit{all four} criteria are met or plausibly attainable. The following chapter applies this utility framework to seven hypothetical neutrino use cases.

%% file: useCaseFindings.tex
\label{useCaseFindings}

The four-criteria \ChapRefFramework{}
offers a common  basis for considering the possible utility of neutrino applications. This chapter applies the framework to seven hypothetical neutrino use cases, each of which received discussion from multiple participants in the Nu Tools engagement sessions. The respective ordering of findings within each category does not represent a prioritization or implied value judgement. 

The capability need and implementation constraints sections present a synthesis of the interview comments, consistent with the concept
that the end-user communities are best equipped to speak to these criteria. The neutrino signature and detection technology sections come from the technology development communities, again following the \ChapRefFramework{}. These sections therefore draw more directly from the knowledge of the Nu Tools Executive Group. For conciseness, the sections below present only considerations specific to each use case that go beyond the general physics of neutrino emission and detection presented in the Nu Tools fact sheets in \autoref{FactSheets}. Readers, especially those new to the field of neutrino applications, may find it helpful to refer to these Fact Sheets as background. 

In brief, three major facts about neutrino physics stand behind the discussion below. First, neutrinos are emitted by fission products in a nuclear reactor (and at a lower level from spent fuel) at a rate proportional to the reactor power; they also carry information about the isotopic content of the fuel. Second, neutrinos rarely interact, which allows them to pass through reactor containment buildings and other surrounding material. This property is the key reason that neutrinos are interesting as a fission signature in applications, yet it also introduces detection challenges. Third, neutrinos are emitted isotropically from fission sources, which means their flux falls with an inverse square law with respect to the source-to-detector distance. Together, these facts enable certain possibilities for applications and also give rise to some limitations. A more detailed discussion of neutrino signals and detection technologies, along with a review of several potential nuclear security applications, appears in a recent review\autocite{RMP}.

\section{Current International Atomic Energy Agency (IAEA) Safeguards}
\label{finding:CurrentIAEA}

\input{useCases/currentSafeguards}

\newpage
\section{Advanced Reactors}
\label{finding:AdvancedReactors}

\input{useCases/advancedReactors}

\newpage
\section{Future Nuclear Deals}
\label{finding:FutureDeals}

\input{useCases/futureDeals}

\newpage
\section{Reactor Operations}
\label{finding:ReactorOperations}

\input{useCases/reactorOperations}

\newpage
\section{Non-Cooperative Reactor Monitoring or Discovery}
\label{finding:NonCoop}

\input{useCases/farNoncooperative}

\newpage
\section{Spent Nuclear Fuel}
\label{finding:SpentFuel}

\input{useCases/spentFuel}

\newpage
\section{Post-Accident Response}
\label{finding:PostAccident}

\input{useCases/postIncident}

%% file: useCases/currentSafeguards.tex
\noindent \textbf{For the vast majority of reactors under current IAEA safeguards, the safeguards community is satisfied with the existing toolset and does not see a specific role for neutrinos.} \\

{\it Summary:} For declared reactors, the current safeguards approach largely relies on containment, surveillance, and item accountancy, and no capability gaps have been identified.\footnote{Fast reactors present a special case, but are commercially used only in Russia, which is a nuclear weapons state and thus not a primary safeguards concern.} These approaches suffice because the fuel in these reactors comes in the form of discrete, countable units. Interest in new technologies is guided by operational ease and time savings without a significant increase in cost. Current neutrino detection technologies do not meet these criteria. 

Neutrinos could provide power and/or fuel burn-up measurements; however, at power reactors, these quantities are typically declared by the operator and are rarely directly measured under safeguards. For research reactors, only reactors with a thermal power over 10\,MW are a significant concern for plutonium production.\footnote{See also \url{https://www.iaea.org/sites/default/files/38402082024.pdf}, accessed June 22, 2021.} There is a small set of such reactors where power is measured by the \IAEA,~in part through thermohydraulic techniques. Neutrino detectors could perform the same task, but it is unclear that they would provide any benefits over existing tools, particularly given the current cost comparison. In summary, neutrino detectors offer capabilities which are either a more expensive duplicate of existing capabilities, such as power measurements at research reactors, or which have no established role and are not seen as a need in the current \IAEA\ practices, such as {\it in situ} burn-up measurements.

\begin{itemize}

\item \textbf{Capability need:}

Most members of the international community, including the US, maintain a policy interest in safeguarding civilian reactors. The goal is to verify that power and research reactors, especially in non-nuclear weapons states, are not used to produce fissile material for weapons programs. This policy is codified in the Treaty on the Non-Proliferation of Nuclear Weapons (NPT), which entered into force in 1970, and is implemented by the International Atomic Energy Agency (IAEA).

Existing reactor safeguards are designed around preventing the diversion of a specific minimum quantity of fissile material, termed a significant quantity (SQ, {\it e.g.} 8\,kg Pu, 25\,kg HEU).\footnote{\url{https://www.iaea.org/sites/default/files/iaea_safeguards_glossary.pdf}, accessed February 4, 2021.}  Power reactors have a relatively limited and consistent set of operational activities, {\it e.g.} receipt and storage of fresh fuel, loading of fuel into the reactor core, transfer of spent fuel to wet storage, and  transfer to dry storage.  These well-defined activities allow, under standard practice, for safeguarding to be accomplished via item accountancy, {\it i.e.}\ counting material, typically in terms of discrete fuel assembly components, as it enters a reactor and comparing against what is removed. Diversion during the intermediate period is addressed by establishing a continuity of knowledge via seals, camera systems, and other technologies.\footnote{\label{footnote1}\url{https://www-pub.iaea.org/MTCD/Publications/PDF/nvs1_web.pdf}, accessed February 4, 2021.} Such mature verification activities are part of a  relatively-static paradigm, with evolutionary technology improvements implemented as appropriate.

In the context of item accountancy, the IAEA relies heavily on its well-established toolset. All safeguards experts interviewed generally felt that existing technologies and procedures are sufficient, and no capability gaps were identified. New technologies could provide capabilities beyond those utilized for item accountancy, such as near real-time measurement of reactor power or fuel burn-up.  However, given that interviewees felt that the item accountancy approach is sufficient, no interviewees saw a benefit to these additional capabilities, particularly when implementation constraints such as costs are considered. New technologies could also play a role in validating fuel-assembly integrity, but cost and signal concerns are significant. 

Research reactors present special challenges for safeguards.  There are currently 222 operating IAEA-classified research reactors worldwide, with 40 that have thermal powers greater than 10~MW.\footnote{\url{https://www.iaea.org/resources/databases/research-reactor-database-rrdb} accessed February 4, 2021.} These reactors have a variety of designs and both flexible missions and operating modes. Beyond preventing diversion of fissile material, safeguarding these reactors for declared activities ({\it e.g.}\ the production of medical isotopes) includes preventing operation of the reactor in ways inconsistent with declared activities to develop capabilities for the production of fissile material. Reactor power monitoring is used to verify declared operation, and in some instances the IAEA uses thermohydraulic power measurements for this purpose.$^{\ref{footnote1}}$
Indeed, there are documented cases of research reactor misuse that were not detected by the IAEA safeguards program.\footnote{\url{https://www.pnnl.gov/main/publications/external/technical_reports/PNNL-25885.pdf}, accessed February 4, 2021.}  Nonetheless, interviewees generally indicated that this reactor class was not a priority for the IAEA.  This was felt to be in part because of the relatively small number of reactors at the upper end of the power range that are suitable for the production of significant quantities of plutonium.  However, neither the importance of this capability gap, nor the technical requirements to fill it, have been well articulated or agreed to by the safeguards community.

\item \textbf{Neutrino signature:}

Neutrino signatures from operating nuclear reactors have been measured over many decades in a variety of basic science experiments. Possible signatures for safeguards have been discussed in a number of papers catalogued in a recent review.\autocite{RMP}$^,$\footnote{See also \autoref{FactSheets}.} The effect of fuel composition is a softening of the neutrino spectrum and reduction of rate with increasing fission contribution from plutonium-239. This effect was experimentally observed as early as 1994\autocite{Klimov:1994}  and most recently has been used to provide a measurement of the uranium-235 and plutonium-239 fission neutrino spectra\autocite{Adey:2019ywk}.

\item \textbf{Detection technology:}

All currently demonstrated reactor neutrino technologies use organic scintillators (which may either be liquid or solid plastic) coupled to photosensors such as photomultiplier tubes.  The detector must produce sufficiently bright and fast (nanosecond scale) scintillation pulses to identify the signal using the two time-ordered pulses from the positron and the neutron. Given the nature of this coincidence signal and the low energies of the deposited radiation, the method is subject to backgrounds from radioactive contamination as well as cosmic rays.

A representative cost for a demonstration detector designed to operate within 10s of meters of a reactor is currently in the $\sim$\$1M--5M range depending on size; costs could reasonably be expected to come down with commercialization. No detector developed to date has been designed for fully remote operation, but there are no significant engineering barriers to doing so. Ton-scale detectors have been constructed within approximately a year for research applications, but this could likely be reduced to several months for a standardized operational design.  Conceptually, ton-scale detectors that could be rapidly deployed onsite as needed are feasible. %

\item \textbf{Implementation constraints:}

Interviewees generally consider the use of neutrino detection in current IAEA safeguards practice at existing reactor facilities to face significant implementation challenges. As noted in the \CCFRefTechnicalReadiness{}, incorporating any new technology into safeguards practice is a large and challenging task. Many interviewees noted this in a general sense, as well as specifically in the context of current IAEA safeguards practice at existing reactor facilities. It was noted that, for a variety of cultural and practical reasons, IAEA is focused on execution. As such, incorporation of new technologies and monitoring concepts, absent a strong capability need, is not an institutional priority. Furthermore, changes in the conceptual approach to monitoring carry a large training and implementation burden, including the possibility that state-level safeguards agreements would need to be amended.

Additionally, multiple interviewees noted that the IAEA has a highly constrained budget, 
and the present equipment outlay for reactor facilities is of order \$100k per site over a five year cycle. Since the capital cost of a neutrino detection system is likely to be at least an order of magnitude higher, these would need to be supported by an external party or the monitored state. While there is precedent for equipment to be provided  in this way, the  holistic lifecycle cost of a neutrino system, including maintenance, operation, and personnel training, would typically fall to the IAEA. For neutrino monitoring to be cost effective in the context of current safeguards needs and practice, it would have to enable the replacement of existing verification technology and, through continuous monitoring, result in demonstrated savings by alleviating the need for onsite physical inspection, thus reducing personnel costs. 

Finally, several interviewees with policy backgrounds consider neutrino monitoring to require further conceptual development. Questions remain about the sensitivity and reproducibility of neutrino detection and how measurements would be interpreted and used to reach safeguards conclusions. Ease of use and reliability were also raised as concerns that would require considerable systems engineering to address.

\end{itemize}

%% file: useCases/advancedReactors.tex
\noindent \textbf{Advanced reactors present novel safeguards challenges which represent possible use cases for neutrino monitoring.} \\

{\it Summary:} In contrast to the case of existing reactors, interviewees expressed a need for alternate technical methods to ensure adequate safeguards of some advanced reactor systems and an interest in the potential role of neutrino detectors to meet this need. This represents an R\&D opportunity as neutrino detector systems can be developed largely independently of the details of the reactor design. Alternatively, the concept of integrating a neutrino detector in the reactor system design may be attractive\footnote{See scenario (1) in the \CCFRefTechnicalReadiness{}.}. 
Overall, neutrino technologies display potential to enable safeguards for advanced reactors where conventional measures such as item accountancy no longer apply. Detailed studies are needed to understand how these technologies would measure parameters of interest.

\begin{itemize}

\item \textbf{Capability need:}

The Department of Energy has expressed a desire to develop the next generation of advanced reactors. These systems must address a variety of challenges, including proliferation concerns. Interviewees report that safeguards approaches for advanced reactors have not yet been fully developed. The issue is timely, given a DOE investment of over \$500 million in FY20 alone\footnote{According to the DOE FY2021 Congressional Budget request.} in support of novel reactors, including \$160 million for new reactor demonstration projects,\footnote{\url{https://www.energy.gov/ne/articles/us-department-energy-announces-160-million-first-awards-under-advanced-reactor}} and the recent \NRC\ Final Safety Evaluation Report for a small modular nuclear reactor (SMR).\footnote{\url{https://www.nrc.gov/reading-rm/doc-collections/news/2020/20-043.pdf} accessed February 4, 2021.}

The advanced reactor space encompasses a wide range of technologies and power levels, including micro-reactors, small modular reactors, molten salt-fueled reactors (MSRs), molten salt-cooled reactors, high temperature pebble bed reactors, and traveling wave reactors.\footnote{See also \url{https://aris.iaea.org/}.} 
Each of these advanced reactor types deviates in relevant ways from the existing fleet of power reactors currently under international safeguards. Traditional material control and accountability methods may not apply, since these methods depend on countable fuel elements, transparent coolant, and frequent refueling. However, any new safeguards solution needs to cater to reactor design features. For example, it would be challenging to verify a MSR via a bulk measurement, due to temporal and spatial variations in its fission rates and fissile content during operations.  This is  especially true if the fuel salts are continuously fed and removed from the core. In contrast, many SMRs will use traditional solid fuel, but they will be part of a complex with multiple modules with more frequent refueling operations.  This operations mode will create a desire for fuel inventory verification without increased inspection resources.

\item \textbf{Neutrino signature:}  

Advanced reactors, depending on their design, can use fast neutron fission, have significantly larger breeding ratios and/or burn higher actinides. Little is know empirically about the precise neutrino yields per fission in those cases. Theoretical calculations indicate that signal rates and proliferation signatures would follow similar trends as for current reactors, with possible exception of increased plutonium breeding in some cases. 
Other distinctive design features of advanced reactors, like non-itemized fuel, more or less frequent refueling, and non-transparent coolants do not directly affect neutrino signal generation or propagation, but they could potentially shift the balance toward new  relevant safeguards capabilities. Nonetheless, some interviewees reaffirmed that the target sensitivity of one significant quantity (1\,SQ) of fissile material remains relevant. 

\item \textbf{Detection technology:}

Since the design of neutrino detectors is not highly sensitive to the details of the nuclear reactor core design, it is expected that the current detector concepts could also be applicable to advanced reactors. There are several specific considerations that apply to advanced reactor monitoring.
Monitoring of bulk fuel may require high sensitivity for neutrino detection, which implies high statistics,  low backgrounds,   and the ability to measure the spectral evolution. 
 For an SMR site with multiple modules, the role of a neutrino-based system may be to verify the power history of individual modules. This could be accomplished using known reactor module-to-detector standoff distances and falloff of the neutrino signal with the inverse square of  standoff distance. In particular, the use of two or more detectors would facilitate module identification via triangulation as has been demonstrated by multi-detector neutrino experiments at multi-reactor facilities\autocite{Abe:2012dc,An:2012db,Ahn:2012nd}. 
Studies show that the breeding blankets in advanced reactors may be monitored via coherent elastic neutrino nucleus scattering (CEvNS) detectors in the future\autocite{Cogswell2016}, but the CEvNS technology has not reached the performance needed to monitor nuclear reactors to date.

\item \textbf{Implementation constraints:}

It was noted by many interviewees that advanced reactors are an emerging technology, and developers are for the most part currently focused on the safety and security of their designs. As such, implementation of safeguards methods is not fully understood and specific requirements for potential neutrino-based methods have yet to be determined. As noted in the capability need section above, a safeguards method and therefore its deployment can vary depending on the reactor design, not all of which can be covered here. Nonetheless, there were several general observations about the potential implementation of neutrino detectors to advanced reactors systems that were noted by interviewees, many of which are common to other findings in this report. 

The cost of neutrino systems, relative to conventional safeguards approaches, was a common concern amongst interviewees. The \UCFRefCurrentIAEA{} provides a full description of such concerns relative to existing safeguards practice and budgetary constraints. However, some important differences for advanced reactors were noted by several interviewees. First, safeguards techniques have not been fully developed for all proposed advanced reactor types, so comparing development, implementation and operational costs is difficult. Neutrino-based approaches may be cost competitive if considered early in the development cycle, where they would not be displacing installed equipment and established procedures; this concept is typically referred to as safeguards-by-design\autocite{sbd}. Additionally, other considerations for reactor developers may make the adoption of neutrino-based approaches attractive. For example, safeguards concerns and associated export regulations may differ for sales to utilities in the U.S.\ vs.\ other nations. The inclusion of additional safeguards systems to provide additional transparency and reassurance of an operator's intent, like one based on neutrinos, might be advantageous or even necessary to allow advanced reactors to be marketed internationally.

Similar to other findings, concerns were also expressed over the physical implementation of neutrino systems with the primary considerations being size and safety. Once again, the possibility to consider neutrino systems relatively early in the reactor development cycle is a difference compared to other findings. The same consideration applies to questions about technical readiness, where addressing requirements for safeguarding advanced reactors could be an integral part of a future neutrino system demonstration. 
It was noted that the similarity of the neutrino signal across the diverse range of advanced reactor types was an attractive feature, since a single system design could have broad applicability.

\end{itemize}

%% file: useCases/futureDeals.tex
\noindent \textbf{There is interest in the policy community in neutrino detection as a possible element of future nuclear deals involving cooperative reactor monitoring or verifying the absence of reactor operations.} \\

{\it Summary:} Neutrino detection is viewed by some experts as warranting further consideration in the context of new treaties and agreements, especially those involving a small number of countries. Verification of new agreements may require novel capabilities. Constraints on verification approaches in a new treaty involving relatively few parties may be less rigid and more open to negotiation compared to those of a large established treaty like the \NPT. 
Agreements with relatively limited scope and a foundation in mutual confidence building are more amenable to the introduction of novel verification technologies. Examples that have been studied previously and/or mentioned during engagements include: fuel disposition treaties in which neutrino detectors could verify fuel burn-up and/or the absence of a plutonium-breeding blanket in a plutonium-burning reactor; a future nuclear agreement in which neutrino detectors could be an area of civil nuclear cooperation, and/or monitoring any future reactor operations; a possible nuclear material treaty in which neutrino detectors could verify core isotopics of reactors fueld with high enriched uranium. Accessing information relevant to these agreements would likely call for a facility-specific deployment. Furthermore, the possibility of combining verification functions with scientific cooperation between participating nations was highlighted as an attractive possibility by some experts.

Nonetheless, as highlighted in the \CCFRefTechnicalReadiness{}, adding to the negotiating diplomats' ``tool box'' depends on a clear understanding of capabilities and deployment requirements. In the case of the recent U.S.-Russia \PMDA and \JCPOA, technical verification techniques already established within the \IAEA\ have served as the default due to their well-understood capabilities and implementation pathways. 
Further work, up to and including a system-level demonstration, is required for neutrino detection to reach sufficient technical readiness to be considered for future agreements.

\begin{itemize}

\item \textbf{Capability need:}

Interviewees expect that the U.S. will continue to seek agreements 
regarding nonproliferation and arms control, building on the NPT and past negotiations in particular regions. Some of these new agreements may cover plutonium-production reactors, among other weapons production facilities. 
 Any new agreements pertaining to these activities would almost certainly include technical verification measures. Interviewees expressed interest in new verification options that would expand the toolset of future negotiators.

In particular, several interviewees expressed interest in new technologies that could verify the shutdown of a plutonium production reactor or monitor the status of a reactor of interest. Currently available approaches include satellite imaging of site activity and heat emission as well as onsite inspections by the IAEA.\footnote{See the \UCFRefCurrentIAEA{} for a discussion of existing onsite reactor monitoring techniques.} A technology combining the precision of onsite inspections and the non-intrusiveness and persistence of satellite imaging would represent a new ``tool in the toolbox” for negotiators, provided it met the implementation constraints discussed below. Technical verification measures that could be deployed before or after a comprehensive onsite inspection were noted as potentially valuable. Some interviewees expressed interest in a capability to exclude the presence of an underground reactor, which may lack a visible heat signature, at a site of interest.  

Beyond verification needs, many interviewees discussed the value of scientific exchange or technical cooperation as a component of future nuclear agreements. Cooperative technical projects in the JCPOA and past U.S.-Russia arms control efforts were cited as historical precedents.\autocite{jcpoa_annex3}$^,$ \autocite{jve} Interviewees indicated that this type of engagement can increase the transparency of verification efforts and present pathways to redirect scientists from a weapons program to other work. In some countries, building general scientific and safeguards capacity may also be seen as a benefit of nuclear agreements with cooperative verification projects.

\item \textbf{Neutrino signal:}

Neutrinos are produced in proportion to the number of fissions in a reactor, and their energy spectrum depends on the fissile material mixture. Neutrinos propagate in straight lines and cannot be shielded. For this reason, the neutrino signal is no different whether a reactor is above or below ground; only the distance from reactor to detector matters, with flux dropping as the inverse square of the distance. The detail that can be extracted from neutrino measurements about a reactor depends on the total number of detected neutrinos, and thus, for a reasonable size detector, decreases with increasing standoff. This decrease is accelerated by the presence of backgrounds from various sources. 

Three neutrino detector deployment ranges can be distinguished in terms of the information they are likely to provide: outside of the reactor building but inside the facility perimeter, at a specific site but outside of the facility perimeter, and the monitoring of any facility within a region. In the first, outside of the reactor building but inside the facility perimeter ($\sim$100~m, $\sim$1~ton-10~ton detector), a neutrino system can deliver a measurement of reactor power, fissile core content, fuel burn-up and potentially fuel enrichment~\autocite{Christensen:2013eza,Christensen:2014pva}. Furthermore, a neutrino system would maintain or enable recovery of continuity of knowledge should it be lost. For MOX fuel, a distinction between weapons-grade and reactor-grade plutonium appears feasible~\autocite{Erickson:2016sdm,Jaffke:2016xdt}. For breeder reactors, the use of coherent elastic neutrino nucleus scattering (CEvNS) may in the future allow the detection of a breeding blanket (for both uranium and thorium fuel cycles).~\autocite{Cogswell2016} In the second range,
at a specific facility but outside of the facility perimeter ($\sim$1,000~m, $\sim$10~ton-100~ton 
detector), a measurement of reactor power and thus a limit on plutonium production is possible. Conversely, a shutdown of a reactor can be verified~\autocite{Carr:2018tak}. At those standoffs, an underground deployment will become necessary to suppress cosmic ray induced backgrounds to an acceptable level. In the third case, monitoring of any facility within a region ($\sim$10,000~m, $\sim$100~ton-1000~ton detector), the presence of any reactor operation above a certain power level can be detected or excluded~\autocite{RMP}. These ranges roughly correspond to differing levels of intrusiveness or access.\footnote{See the \CCFRefSiting{}.}     

\item \textbf{Detection technology:}

As noted above, technology selection for use in a future nuclear deal will depend on many factors, including the information needed for verification and implementation constraints like cost and allowable deployment site. A number of neutrino physics experiments and detection demonstrations provide real-world examples of possible technology options.

For use cases requiring precise neutrino rate and/or spectrum information, siting close to the reactor will be necessary. In the fortunate event of a shallow underground site being available at such close range, ton-scale scintillator detectors without significant internal segmentation have demonstrated good performance\autocite{Bowden:2006hu,Boireau:2015dda,Ko:2016owz}. In the more likely event of a surface site being the only option, ton-scale segmented scintillator systems that incorporate a neutron capture agent for inverse beta decay identification have recently demonstrated neutrino detection\autocite{Ashenfelter:2018iov,Haghighat:2018mve}.  

Should a use case require detector siting at greater standoff, underground deployment becomes necessary to suppress background. At distances of $\sim$100~m-1,000~m, three experiments have demonstrated high performance single volume scintillator detectors at the 10~ton-20~ton scale\autocite{Abe:2012dc,An:2012db,Ahn:2012nd}. When operated $\sim$50~m-100~m underground these systems had good sensitivity, being able to measure rate and spectrum a few 100~m from large power reactor complexes.

Several efforts have demonstrated or plan larger detectors that provide sensitivity to reactor neutrinos at greater distances. These $\sim$1000~ton systems must be operated deep underground (100~m or more) and use scintillator\autocite{Eguchi:2002dm}, water doped with a neutron capture agent\autocite{FERNANDEZ2016353,AIT}, or water-based liquid scintilator (WbLS).\autocite{askins2020theia}

CEvNS of reactor neutrinos has been suggested as an alternate detection method, specifically because it is sensitive to low energies that are invisible to inverse beta decay. While no system that can identify the low energy signal from this reaction over background has yet been demonstrated, research and development on several technologies are underway in a basic science context. 
 
\item \textbf{Implementation constraints:}

The inclusion of neutrino technology in any future nuclear deal will require that a demonstrated solution be well-established in the  
negotiators' toolset. Such a solution would be viable for consideration only after a high degree of field testing and a performance record that exhibits a false-positive rate commensurate with the stakes of maintaining compliance. The technology development costs to meet this threshold are significant
and would likely be borne by nations rather than international institutions.
Additional concerns about operation and maintenance costs remain a significant impediment for the consideration of neutrino detectors.

Many of the reasons neutrino solutions are attractive are also the sources of new challenges and concerns to participants of future nuclear deals. A major concern is political acceptability given the novelty of the technology. 
Participants may have concerns that a neutrino instrument could be revealing information on activities beyond those of the scope of the agreement. Such concerns and the role of open scientific collaborations would have to be covered by future agreements.

Considerations discussed in the \CCFRefSiting{} favor system siting as close as reasonably achievable to an individual reactor facility of interest. Concepts for exclusion or aggregate monitoring by a single detector of reactor operations at known facilities over a local area with linear dimensions $\sim$10~km may shift these considerations. 
Such an approach would have to be motivated by political factors such as mutual confidence building. 
While a handful of interviewees expressed interest in such concepts, concerns over large costs were common.

\end{itemize}

%% file: useCases/reactorOperations.tex
\noindent \textbf{Utility of neutrino detectors as a component of instrumentation and control systems at existing reactors would be limited.} \\

{\it Summary:} Experience with current reactor designs spans nearly eight decades and has yielded a mature suite of instrumentation to guide operators, both in safety-related functions and overall performance monitoring.  Neutrino detection rates imply that the time required to determine a change in reactor state is long compared to what is needed for safety-critical instrumentation. For these reasons, no role is seen for neutrinos as part of the operational infrastructure. While neutrinos could provide cross calibration of instrumentation subject to harsh environments, e.g. a re-calibration reference for \BWR{}, no detailed studies of this potential capability have been performed to date.

\begin{itemize}

\item \textbf{Capability need:}

Safe and efficient reactor operations require knowledge of the reactor power, core neutron flux, and other parameters, on the second to minute timescale. In existing power reactors, these quantities are determined using a complex suite of instrumentation in and outside the reactor core. Instruments used in water-cooled reactors include coolant temperature sensors and neutron detectors with designs that have been optimized over decades of reactor operations. 
In contrast to other findings, the primary end-user is the nuclear industry and not the safeguards community. Interviewees report that the nuclear industry is generally satisfied with the instrumentation available for existing reactors. The ability to maintain continuity of knowledge during a safety event would be valuable,\footnote{See the \UCFRefPostAccident{}.} but no specific scenario relevant to neutrinos and not already covered by existing systems was identified.

\item \textbf{Neutrino signature:}

The typical neutrino signal rate described in \autoref{FactSheets} corresponds to an interaction rate of about 1 interaction per minute per gigawatt in 1 ton of detector at a standoff of 25\,m. That is, a neutrino system will need at least 10s of minutes to respond to a change in reactor conditions and thus, is precluded from being a part of the safety related instrumentation at a reactor. With sufficient integration time, on the order of weeks to months, neutrino measurements of reactor power at the percent level or better seem feasible and thus neutrino signals could be used to absolutely calibrate less accessible instrumentation. 
Further study would be needed to assess the usefulness of this concept.

\item \textbf{Detection technology:}

Neutrino detection technology that could in principle be relevant to reactor operations is similar to that described in the \UCFRefCurrentIAEA{} and the \UCFRefAdvancedReactors{}.

The low neutrino cross section limits the interaction rate and results in a relatively slow response time from existing neutrino detectors. This is seen as too limiting for instrumentation and control applications, which typically require  measurement and control feedback on the second to minute timescale. No path has been identified in neutrino detection technology development that would fundamentally alter this characteristic.

\item \textbf{Implementation constraints:}

As with other findings, cost considerations were often mentioned by interviewees as an implementation concern with respect to the use of neutrino-based instrumentation for existing reactor types. However, in this case there were divergent views, with one interviewee deeming a \$1M-scale detector reasonable if it were to provide economic benefits to operators, while another discussed severe budget constraints for instrumentation. 

The need to maximize signal necessitates locating a system as close as possible to, or within, a facility.  As discussed in other findings, conforming with safety and security regulations at existing facilities thus will yield important implementation constraints.  All currently demonstrated reactor neutrino technologies use combustible organic scintillators, with many also incorporating combustible hydrogenous shielding materials. Standard ignition mitigation techniques like fire-proof skins or blankets have been used in demonstrations to date to address regulatory requirements when deploying within combustible exclusion areas in reactor facilities. Demonstrations have  often used liquid organic scintillators. Use of liquids at the ton-scale requires engineering solutions to comply with spill-control regulations. Modern high-flash point liquid scintillators have largely addressed flammability concerns, but use of liquid still carries a negative perception with some interviewees. Solid plastic-based systems would address such concerns and ease deployability, but to date have not matched the performance of liquid-based systems. Another regulatory concern results from the complexity of neutrino instruments.

As noted for the detection technology criterion, the limited interaction rate of neutrinos would preclude their use for safety critical instrumentation. This consideration also involves implementation constraints since the achievable detection rate is strongly related to system location and size. Since implementation of a neutrino-based system will have to adapt to the layout of an existing facility, these system parameters will be constrained, limiting flexibility in the design process if trying to meet a detection rate requirement. Lastly, multiple interviewees noted that the use of neutrino-based instrumentation at multi-reactor facilities would require techniques to disambiguate the signal from each core. Further conceptual development of system implementation would be required to address comments of this nature, in addition to detector R\&D.

\end{itemize}

%% file: useCases/farNoncooperative.tex
\noindent \textbf{Implementation constraints related to required detector size, dwell time, distance, and backgrounds preclude consideration of neutrino detectors for non-cooperative reactor monitoring or discovery.}\\

{\it Summary:} In principle, neutrino detectors can monitor known or discover clandestine reactors from beyond the borders of a country of interest
without the cooperation of that country’s government. Some experts find these intelligence-gathering capabilities attractive as a part of national technical means, in order to monitor states that do not grant access to international inspectors. Others believe there is no real capability need, given the coverage of existing national technical means, chiefly satellites. 
Experts agree that the practical challenges of long-range non-cooperative reactor neutrino monitoring outweigh whatever hypothetical benefit they may provide.

The key constraint for far-field monitoring is detector size. Because neutrino signal strength drops with the square of standoff distance, very large detectors are required to obtain sufficient neutrino signal for reactor detection at long range. 
Large detector size confronts multiple severe constraints in the end-user community, including construction and operation costs, construction timeline, and the potential concerns associated with a country conspicuously monitoring its neighbor.
Together, these practical constraints preclude any realistic possibility of non-cooperative,  
long-range reactor monitoring.
Close-range monitoring, with either a mobile or stationary system, would require cooperation of the host country because the required detector mass and dwell time are not compatible with a covert operation.

\begin{itemize}

\item \textbf{Capability need:}

The capability to observe nuclear facility operations over large regions without local facility cooperation is an attractive intelligence capability to many survey interviewees. Measurement capabilities that can augment facility and reactor declarations of activity through all phases of a project can be valuable. In particular, the detection of small-scale, intentionally-hidden production reactors is often discussed as a highly-motivated capability target for neutrino detectors. Knowledge of such programs is generally obtained through flexible monitoring methods (such as satellites) that can observe a variety of activities associated with the build up to an operational nuclear program. That is, unlike neutrino detectors, they have the potential to provide necessary information well prior to a reactor becoming operational. 
\item \textbf{Neutrino signature:}

Non-cooperative reactor monitoring faces a fundamental constraint from the weakness of neutrino interactions.

One deployment mode possible without the
cooperation by the host country is siting outside of the country, 
implying standoff distances of 100s of kilometers or more. The neutrino signal drops like the inverse squared distance, so at 100\,km distance the signal rate will be 1/10,000 of what it is at 1\,km. Therefore, detector size goes from the 10s of ton range into the 100s of kiloton (100,000s of ton) range even in the absence of backgrounds.

On top of the fundamental limit imposed by the weakly interacting signal, backgrounds further increase the detector size and deployment complexity for non-cooperative concepts. 
For a detector at 100 m standoff, the presence of even moderate backgrounds\footnote{\textit{i.e.} the signal-to-background ratio of demonstrated surface detectors.} raises the threshold for a useful signal to 30,000 kg-days. 
For a beyond-border detector,
backgrounds from cosmic rays force deployment deep underground. Eventually, neutrinos from other operating reactors in the region
become a dominant contributor.
The following example illustrates 
this issue: Europe has a total installed reactor power of around  250~GW$_\mathrm{th}$ which will result in the same neutrino signal as one 100~MW$_\mathrm{th}$ reactor at a 50 times larger distance, {\it e.g.} for a standoff of 100~km from the reactor of interest even reactors as far as 5,000~km from Europe will have to contend with the resulting background. 
Reactor backgrounds could be reduced through
determination of the neutrino direction on an event-by-event basis. In inverse beta decay, the momentum, and thus the direction of the neutrino, is carried away by the resulting neutron ($E_n\simeq 50$~keV). Thus, event-by-event reconstruction of the neutrino direction necessitates neutron momentum reconstruction in a 
 large detector, which is not a current, emerging or even hypothetical capability.
In elastic electron-neutrino scattering, the direction information is preserved to a lesser degree, but it is contained entirely in the recoiling electron ($E_e\simeq 4$~MeV), which in principle can be tracked even in a large detector. However, the interaction rate per unit detector mass in water is a factor of 5 lower~\autocite{Barna:2015rza} and depends on the achievable detection threshold.  Here, ultimately intrinsic radioactive backgrounds from the detector itself will be the limiting factor. The information contained in the very weak neutrino signal would at best allow an on/off (presence/absence) declaration or an upper limit on time integrated reactor power.

The other non-cooperative  mode is a close-range covert deployment.
To obtain a handful of events at 100\,m standoff from a 100\,MW reactor, the product of detector mass and dwell time needs to exceed 1,000 kg\,day. 
That is, even a perfectly efficient, zero-background, ton-scale (truck-sized) neutrino detector would need to be deployed within 100 m of the reactor core for a full day to collect a few signal events.

\item \textbf{Detection technology:}

A variety of technologies capable of detecting neutrinos at short and long range have been demonstrated. However, detecting a sufficient signal in a non-cooperative deployment is a challenge due to the fundamental physics constraints laid out above. For a close-range deployment, the demonstrated near-field technology based on segmented scintillator would face the mass and dwell-time issues noted above.
For a  beyond-border deployment,
the low neutrino flux 
requires large detector volumes to achieve a sufficient neutrino interaction rate. 
Possible detection media include water, organic scintillator, and water-based scintillator. 
In water, neutrino interactions are detected via Cherenkov light.
By doping the water with a neutron capture agent such as gadolinium a delayed coincidence signature for IBD events is achieved. Such a  signature combined with  energy and fiducial volume requirements can dramatically reduce backgrounds; nevertheless, 
 surface cosmogenic backgrounds are so large that they must be reduced by deep underground deployment.
An example of the current state of technology able to detect neutrinos from remote reactors is the Super-Kamiokande detector, which is located approximately 1000~m deep and  contains 50~ktons of water in the largest detector tank currently used in a neutrino detector.

\item \textbf{Implementation constraints:}

Non-cooperative use cases involving neutrino detection face severe implementation constraints. These derive from the general consideration of achieving a sufficient signal rate, which as noted above depends on a combination of detector size, background suppression, standoff distance, and dwell time. For non-cooperative operation within the borders of a country, one must consider small, portable detectors, which implies small standoff distance and long dwell time requirements that would be incompatible with a covert deployment.

In the case of cross-border, non-cooperative monitoring from the territory of a willing host, interviewees generally agreed that the use of neutrino detection would be impractical, primarily due to the large detector size that would be required. Non-cooperative monitoring or verification taking place within the borders of a neighboring country implies a large standoff distance, in turn leading to multi-kiloton scale detectors deployed underground to provide sufficient overburden. A construction project of this scale would be difficult to execute without drawing the attention of the country being monitored, raising questions with respect to a potential escalation of regional tensions.  Extended construction projects also provide an opportunity to reconsider the location and operation of reactor facilities. Another implementation consideration arises from the relatively limited fraction of a country's geographic area that could be monitored in most cases. 
Information as to the suspected location of an undeclared facility would be needed when choosing the construction site for a large underground neutrino detection system. Alternately, a known facility that one wished to non-cooperatively monitor for verification purposes would have to be located conveniently close to a border. Additionally, defining legal mechanisms to facilitate non-cooperative activities  would be complex, {\it e.g.\ }the host country would need to agree to construction and surveillance activities over a long time period.

Further considering cross-border non-cooperative monitoring at long distances, the system cost for a sufficiently large detector and the required underground cavern represent another severe implementation constraint. While some interviewees expressed the view that long distance detection capabilities could in principle be useful, none expressed the view that the utility provided by foreseeable technological implementations justified the associated cost scale.

\end{itemize}

%% file: useCases/spentFuel.tex
\noindent \textbf{Non-destructive assay of dry casks is a capability need which could potentially be met by neutrino technology, whereas long-term geological repositories are unlikely to present a use case.}\\

{\it Summary:} The inventory of spent nuclear fuel in dry cask storage continues to increase globally due to a lack of operational permanent geological repositories in most countries. 
The current approach to maintain a continuity of knowledge (CoK) with respect to cask contents is based on seals and video surveillance.  
Interviewees report that no suitable non-destructive assay technology exists despite efforts to develop this capability based on neutron and gamma signatures. Neutrino signatures contain sufficient information for this task, but the neutrino emission from spent fuel is orders of magnitude lower than from an operating reactor, creating strong challenges for detection.

Permanent storage of spent nuclear fuel in geological repositories is still under development, and only Finland and Sweden plan to start operations of such a facility within this decade. There is no consensus in the safeguards community on what level of assurance is needed for these facilities after closure, but in general the IAEA requires dual Containment and Surveillance (C\&S) with no common failure modes. 
In principle, neutrino detectors could supplement C\&S safeguards for a geological repository, but the low neutrino signal would require such large detectors that it appears unlikely relevant capabilities could be obtained in practice.

In summary, non-destructive analysis of dry cask contents is an unmet capability need and relevant neutrino signatures for this task exist, although the detection challenges are formidable. 
In the case of geological repositories, the detection and implementation challenges likely outweigh the possible monitoring value.

\begin{itemize}

\item \textbf{Capability need:}

The safeguarding of spent fuel in dry casks requires  maintaining CoK following transport out of the reactor building. In this case, containment and surveillance, especially tags and seals, are used to ensure CoK. As inspection or replacing these seals requires  access to the top of casks, there is desire to reduce the physical and radiation risks to inspectors. The high risk and resource requirements associated with opening  a previously closed cask gives great importance to visual inspection of the tags and seals. Non-destructive assay (NDA) methods are therefore desired to re-verify cask contents and reduce demands on inspection resources. NDA techniques based on neutrons and gammas have little to no sensitivity to potential diversions of material from the center of the cask due to self-shielding. Any proposed technology should have high reliability, longevity, and  provide verification independent from declarations. The insufficient performance of neutron and gamma NDA systems has motivated the exploration of new technologies such as muon tomography. These approaches have not yet yielded acceptable solutions, and thus there remains a clear need for additional capabilities.

There is no agreed upon general concept of safeguards for geological repositories. The main concern is the content of fissile material. Sweden and Finland will start operations of final repositories within the decade, and it appears that in these cases no specific arrangements for safeguards have been established.  Containment and surveillance methods are anticipated to be used, although additional monitoring capabilities are desired. Verification of spent fuel inventory requires technology with minimal maintenance needs and the ability to operate over a long timescale. Technologies that have the dual function to ensure environmental and criticality safety of long-term disposal are valued.

\item \textbf{Neutrino signature:}

The neutrino signal from spent nuclear fuel arises from the beta decay of fission fragments.\footnote{Beta decays from actinides provide a negligible flux of neutrinos compared to fission fragments.} The rate of neutrino emission from spent fuel is much lower than from an operating reactor because most fission fragments produced in a reactor have short half-lives, on the order of seconds to minutes. As soon as fission ceases, the neutrino signal begins to drop precipitously. Five minutes after fission stops, the emission has dropped by about a factor of 10, after a week it drops by another factor of 10, with continued decreases over the subsequent months and years: after 10 years the rate is down to $10^{-5}$ relative to a fissioning system\autocite{Brdar:2016swo}. 
The only fission fragment producing a significant number of neutrinos above the threshold for inverse beta decay 
over periods longer than a decade is strontium-90. It has a high (percent level)  yield in the fission of both uranium-235 and plutonium-239 and a half-life of 29 years. To date, neutrino detectors have not demonstrated the ability to detect neutrino emissions from spent nuclear fuel. However, geoneutrinos which fall in the same energy range and present a very low intensity neutrino signal have been detected by two independent experiments\autocite{Araki:2005qa,Bellini:2010hy}. Simulations have predicted that the signal in a 20-ton detector placed within 50~meters of a dry cask storage facility is statistically strong enough within one year to observe the removal of spent fuel from 1 spent fuel cask \autocite{Brdar:2016swo}, not accounting for the likely very sizeable backgrounds close to the surface.

\item \textbf{Detection technology:}

The design and construction of detectors for spent fuel will be generally similar to other reactor neutrino detectors. Deployment mode, size and background issues would, however, differ between the spent fuel cask dry storage and geological repository applications. In the case of a dry cask storage facility, the detector would be located above ground or at a shallow underground site (meters of overburden) to either track the movement of fuel casks or to re-verify the contents of a single cask. In this application detector masses of 10-50\,ton are appropriate and the challenge is to reduce backgrounds to an acceptable level beyond what has been achieved experimentally\autocite{Ashenfelter:2018iov}. Current operating detectors suggest a segmented design is required. Directional reconstruction of neutrino events would be desirable to reduce backgrounds and increase the sensitivity.

For kilometer-scale geological repositories, the detector would be deployed deep underground, on the order of a few hundred meters from the facility, and have a mass in the 100-10,000~ton range. Here, backgrounds would not be an issue and existing neutrino detectors like KamLAND\autocite{Eguchi:2002dm} are suitable; however, such a detector  would have  little sensitivity to anything other than cataclysmic events that impact the entire facility. In order to have statistical sensitivity to  anomalies at a relevant level on required timescales, a detector would need to have the capability of crude imaging of the facility. This requires an angular resolution of the order 10 degrees\footnote{Here directionality is primarily needed to improve sensitivity and not to deal with backgrounds.}, which can only be achieved by using electron-neutrino scattering. This implies an increase in the detector size to several kilotons. The resulting combination of attributes, large mass and angular resolution,
 is currently unavailable and difficult to foresee for the future in scintillator detectors. A recent study looks into the use of neutrino-electron scattering in conjunction with a liquid argon time projection chamber\autocite{Goettsche2020}. This is a detector technology which could be directionally sensitive and  is being developed at a scale of  40,000~tons for the DUNE experiment\autocite{Abi:2020loh}.

\item \textbf{Implementation constraints:}

The implementation implications of the low signal rate and associated measurement timeline for relatively weak spent fuel sources yielded mixed responses from interviewees. One interviewee expressed concerns about the sensitivity to storage containers with older fuels, while another interviewee discussed the low rate in which casks are received in some facilities, which would tolerate a longer detection time. Further concerns and requirements regarding sensitivity were also expressed. For neutrino detectors to replace existing verification technology, they need to meet high sensitivity standards, such as detecting a discrepancy of one significant quantity of material in certain scenarios or verifying a 50~metric-ton inventory of fuel to 10\% precision. 
For fuel storage applications, concern was expressed that using neutrino detectors would require changes in standard procedures, in particular moving fuel casks to an emplaced detector system. 

Similar to other findings, concerns were expressed about the general timeline associated with adopting new technologies, specifically by the IAEA. For domestic safeguards of fuel storage and geological repositories experimental validation would be required by an industrial partner who manufactures dry casks, as well as certification of use for periods of time counted in decades.  Neither liquid scintillator nor liquid argon based neutrino detectors have  been operated for these timescales; while operating they do require access for maintenance.   
In addition to the time to implement and validate a new technology, existing facilities would need to be retrofitted to house additional instrumentation. It was pointed out that requirements for robustness, longevity, and environmental safety will be more stringent for geological repositories that are located below the water table.

As noted in multiple findings, cost is a concern for implementing neutrino detection. For fuel storage 
repositories, there were divergent viewpoints on whether neutrino system cost would be reasonable or too high. 
In the context of fuel storage applications, it was noted that the cost of conventional nuclear detectors, such as multiplicity counters, can approach \$1M, but also that the low signal strength may require large and therefore even more costly neutrino detectors. As in other Findings, multiple interviewees discussed the limited IAEA safeguards budget and their inability to support procurement of items in this cost range. 

\end{itemize}

%% file: useCases/postIncident.tex
\noindent \textbf{Determining the status of core assemblies and spent fuel is a capability need for post-accident response, but the applicability of neutrino detectors to these applications requires further study.} \\

{\it Summary:} During the nuclear power accidents at Three Mile Island, Chernobyl, and Fukushima Daiichi, the initial response aiming to stabilize the facilities and protect the public was hampered by significant information gaps. In particular, responders lacked continuous knowledge of the fuel’s physical location, configuration, and criticality status long after the accident.  Multiple interviewees pointed out that, despite considerable investment, there remains a need for instruments capable of determining the location and configuration of fissile material in a wide range of post-accident scenarios.

Neutrino detection could in principle play a role in a providing an indicator of ongoing fission reactions.
However, much is currently unknown about the performance requirements a neutrino detector would have to meet for this use case.
Requirements for fission power sensitivity would likely be more stringent than presently demonstrated.
A detector would need to operate in a challenging post-accident environment, including possible radiation fields and limited access to power sources. 

Additionally, it is not clear which stakeholders among governmental agencies, industry associations, reactor vendors, and reactor operators would adopt the responsibility of supporting R\&D of this type. 
Further expert engagement will be required to understand response sensitivity requirements and reasonable parameters defining potential operating environments. 
The technical feasibility of appropriate detection technology could then be assessed.

\begin{itemize}

\item \textbf{Capability need:}

Reactors are designed with a ``defense in depth" strategy that accounts for the potential of multiple simultaneous failure modes to prevent any accident from resulting in radiation release. However, in the case of a reactor-related incident, a portion of an appropriate response requires basic knowledge of the core condition. The three most significant global accidents, Three Mile Island (1979), Chernobyl (1986), and Fukushima Daiichi (2011), identified information gaps in technical parameters associated with the core state following an incident. In cases where core melting might have occurred, it is necessary to know the location, quantity, and configuration of melted fuel. 
Information about the core is needed to understand both the immediate criticality status and the potential for future changes.  While there exists a suite of tools for detecting criticality, {\it e.g.\ }gamma radiation, neutron radiation, and volatile fission daughters, 
in the incident at the Fukushima Daiichi plant significant damage to existing instrumentation and radiation-restricted access limited available data. Radiation levels were high enough in key locations that instrumentation sent in after the event was rendered inoperable.  As a result, even after ten years, there is still a lack of quantitative knowledge of the situation with the cores in units 1, 2, and 3.

Furthermore, the situation can be dynamic; radiation flare-ups identified by instrumentation have occurred well after the Fukushima incident.  It remains a challenge to determine if these result from shifting of bulk material ({\it i.e.\ }changes in shielding) or changes in the criticality of core material. Some interviewees suggested that there could be post-accident scenarios where the condition of spent fuel would be a concern.\footnote{Fast reactors may pose a unique challenge as a moderating material is not needed to sustain the chain reaction.}

\item \textbf{Neutrino signature:}

In a post-accident scenario, the fission rate in the reactor core will likely be much lower than during normal operation. While the neutrino emission rate cannot be predicted in advance since it will strongly depend on the accident scenario, residual neutrino emissions from previously irradiated fuel provide a lower bound.\footnote{See the \UCFRefSpentFuel{}.} A damaged core may present a sizeable neutrino source for days to weeks after fission has ceased. Neutrino propagation from source to detector is unaffected by intervening material and thus could represent an attractive signal in cases of bulk material shifting.

\item \textbf{Detection technology:}

At this time the conditions that define post-accident scenarios have not been studied, and as a result, there is limited understanding of the magnitude of the neutrino signal that would have to be detected. Assessment of the sensitivity of neutrino detection to post-accident conditions and the selection of appropriate technologies await better definition of the signal source term. The harsh radiation environment inherent to many post-accident scenarios is likely to require additional shielding and/or background rejection capabilities.

\item \textbf{Implementation constraints:}

Multiple interviewees commented that a neutrino-based system would need to be compact, transportable, and robust against a possibly harsh radiation environment to meet the implementation constraints of a post-accident nuclear reactor scenario. An alternative implementation is to locate the detector permanently on site, although this strategy would require integration with facility design and may have physical deployment constraints.\footnote{See scenario (1) in the \CCFRefTechnicalReadiness} Although the performance requirements on post-accident diagnostics are as yet poorly defined, interviewees consistently called for autonomous systems, likely with power access, that could be deployed close to the reactor site. Additionally, as noted in the \UCFRefReactorOperations{}, the dwell time is an important implementation consideration given the likelihood of a low signal rate. Post-Accident scenarios can be dynamic, and reliable information, based on a sufficiently large dataset, may be needed on short timescales. This consideration is in obvious tension with the desire for a compact system.

\end{itemize}

%% file: recommendations.tex
\label{recommendations}

On the basis of the study Findings, the Nu Tools Executive Group makes the following pair of recommendations to the study sponsor, \DNNRD. Pursued concurrently, cooperatively, and with equal priority, the recommended actions present a pathway to applying the capabilities of neutrino technology in service of nuclear energy or security needs. The recommendations have equal priority. 
They reflect the \ChapRefUCFindings{}, which indicate that the areas with the most potential utility for neutrino detectors are \hyperref[finding:FutureDeals]{\emph{Future Nuclear Deals}} and \hyperref[finding:AdvancedReactors]{\emph{Advanced Reactors}},
along with possibilities for further study in 
\hyperref[finding:PostAccident]{\emph{Post-Accident Response}} and 
\hyperref[finding:SpentFuel]{\emph{Spent Nuclear Fuel}}.
Consistent with the \ChapRefCCFindings{} and four-part \ChapRefFramework{} developed in the Nu Tools study, the recommended actions drive toward a meeting point between the needs and constraints of the nuclear energy and security communities and the capabilities offered by neutrino physics and technology. The Nu Tools Executive Committee recommends that \DNNRD~pursue both recommendations simultaneously in future investments.

\section{Recommendation for End-User Engagement}
\label{rec:EngagementSupport}

\noindent {\bf DNN should support engagement between neutrino technology developers and end-users in areas where potential utility has been identified.} \\

\DNNRD~should allocate and consistently provide appropriate resources for technology developers and end-users to establish and maintain a dialogue.  This can be accomplished by supporting specifically charged working groups, establishing targeted topical meetings, and supporting attendance of well-established meetings such as the Institute of Nuclear Materials Management Meetings and American Nuclear Society Meetings. \DNNRD~should provide targeted support to conduct modeling and simulation studies that will evaluate potential performance of neutrino detectors in specific use cases, including advanced reactors, future nuclear deals, and the less-developed use cases of post-incident response and spent nuclear fuel monitoring. \DNNRD~should encourage the inclusion of appropriate technical experts outside of the neutrino detection community in these studies to ensure they have high relevance to end-users. The four-criteria framework developed in this study serves as a useful tool to structure the exchange between the technology development and end-user communities; \DNNRD~should encourage its adoption in program planning and evaluation.
\\

\section{Recommendation for Technology Development}
\label{rec:Portfolio}

\noindent {\bf \DNNRD~should lead a coordinated effort among agencies to support a portfolio of neutrino detector system development for areas of potential utility, principally for future nuclear deals and advanced reactors. }\\

A key determination from this study is that communication and coordination across agencies and stakeholders is needed to establish technical approaches that address realistic use cases.
Neutrino system development within \DNNRD{} would be enhanced by drawing on relevant technical and project execution expertise found in communities supported by basic science agencies. Communities with expertise in the most promising utility areas should also be involved in defining specific needs as technology matures and approaches a demonstration stage. \DNNRD\ should take the lead in this coordination, since it is the most significant stakeholder in neutrino technologies for nuclear energy and security applications. 

A coordinated effort led by \DNNRD{} to develop and demonstrate a neutrino system at appropriately high TRL  in an application-relevant context is a necessary step to bridge the gap between R\&D and actual adoption by end-users.
To succeed in this goal, support for neutrino signature prediction and calibration must be included alongside that for detector development. 
Since it takes an extended period of time to transition a technology to higher TRLs, coordination and research funding should proceed 
in a parallel manner while identifying demonstration opportunities relevant to specific areas of potential utility.

Support should be prioritized for technology developments that will enable full exploration of the most promising utility areas identified by this study: future nuclear deals and advanced reactors. \DNNRD\ should increase its  investment in detector technologies which allow for surface deployment of ton-scale detectors.

%% file: glossary.tex
\label{glossary}

\begin{description}
\item[AAP] is an annual workshop series on {\it Applied Antineutrino Physics}.  This meeting is primarily attended by technology developers, but potential end-users are encouraged to participate. 
\item[Antineutrinos] are the antimatter partner to neutrinos.  In plainspoken language it is common to use the word ``neutrinos'' to refer to both neutrinos and antineutrinos unless it is important to stress the difference, as is almost never the case in this report.  Nuclear reactors primarily produce antineutrinos through the beta decay of neutron-rich fission fragments and the inverse beta decay detection process is only sensitive to antineutrinos.
\item[BWR] {\it Boiling Water Reactors} are a common type of power reactor that use low enriched uranium fuel and generate steam in the primary reactor vessel.
\item[Cherenkov light] is produced when a charged particle, such as the positron produced in an IBD interaction, is moving faster than the speed of light in a medium.  Cherenkov light is emitted in a cone about the direction of the charged particle's trajectory.  Water is a commonly used medium for larger neutrino detectors that rely on the Cherenkov process to track the charged particles produced in neutrino interactions.
\item[CEvNS] stands for {\it Coherent Elastic Neutrino} ($\nu$) {\it Nucleus Scattering}, in which a neutrino (or antineutrino) scatters off of a whole nucleus.  This process has a greatly enhanced probability relative to other neutrino interaction processes, including IBD, particularly with heavy nuclei, but the signal is very hard to detect.  CEvNS has only recently been observed using neutrinos with energies that are about 10 times greater than typical reactor neutrinos.  This type of scattering should exist for reactor neutrinos, but it has not yet been observed.  Thus the potential for applications of CEvNS scattering to reactors seems promising, but it is still speculative.    
\item[CoK] {\it Continuity of Knowledge} refers to the system of data that provides uninterrupted information in order to prevent undetected material production/transport or undeclared facility operation.
\item[Cosmic rays] are high-energy particles created when energetic protons or other atomic nuclei strike the upper atmosphere.  These particles rain down on the Earth's surface and create a persistent background in all neutrino detectors.  These backgrounds can be dealt with by using active measures to tag and reject cosmic rays, or by placing detectors under thick shielding to attenuate the cosmic rays.
\item[C\&S] {\it Containment and Surveillance} are safeguards techniques applied to maintain continuity of knowledge (CoK) through verification of nuclear material transfer at decalred points. C\&S technologies typically include optical cameras and seals.
\item[DNN] is NNSA's {\it Office of Defense Nuclear Nonproliferation}.
\item[DNN R\&D] is the {\it Defense Nuclear Nonproliferation Research and Development}, which is the R\&D arm of DNN.  It is also known by its DOE program designation, NA-22.
\item[DPRK] stands for the {\it Democratic People's Republic of Korea}, commonly known as North Korea.
\item[FMCT] is the {\it Fissile Material Cutoff Treaty}, a proposed international treaty to prohibit the further production of fissile material for nuclear weapons or other explosive devices.
\item[Geoneutrinos] are electron antineutrinos from radioactive isotopes in the geological materials, typically rocks.  These antineutrinos primarily come from isotopes in the uranium-238 and thorium-232 decay chains, and from potassium-40. 
\item[GWth] stands for {\it Gigawatts Thermal}.  It is a unit for the total thermal power of a reactor, as opposed to the electrical power, which is typically about one third of the thermal power.   
\item[HEU] stands for {\it Highly-Enriched Uranium}, which corresponds to a uranium-235 concentration of 20\% or greater.
\item[IAEA] is the {\it International Atomic Energy Agency}, an autonomous agency within the United Nations system based in Vienna Austria.  It is responsible for verifying nations' compliance with their obligations under the NPT.  
\item[IBD] stands for {\it Inverse Beta Decay}, in which an electron antineutrino ($\bar{\nu}_e$) exchanges charge with a free proton to become a positron and a neutron.  This is overwhelmingly the most common process for detecting reactor neutrinos, because it has a clean, coincident detection signature comprised of a prompt signal formed by the kinetic energy and annihilation of the positron, followed by a delayed signal from the neutron capture.
\item[JCPOA] is the {\it Joint Comprehensive Plan of Action}, commonly known as the Iran nuclear deal, a 2015 agreement between Iran, the United States, France, Russia, the United Kingdom and Germany on the Iranian nuclear program.
\item[MSRs] {\it Molten Salt Reactors} are a class of advanced reactor that use a liquid salt. They come in two sub-classes, one in which fissile material is dissolved into the salt and another in which the salt serves only as a coolant. MSR designs can utilize a thermal or fast neutron spectrum. They typically operate with higher temperatures and closer to atmospheric pressure when compared to light water reactors. More details can be found in an  overview.\footnote{\url{https://doi.org/10.1016/j.pnucene.2014.02.014}}
\item[m.w.e] stands for {\it meters water-equivalent}.  It is a measure of cosmic-ray shielding.  For an underground detector, typical undisturbed rock has a density of 2.6\,g/cm$^3$, such that 0.38\,m of rock corresponds to 1\,m.w.e shielding. 
\item[MOX] stands for {\it Mixed Oxide} Fuel.  It is nuclear fuel that contains more than one oxide of fissile material, usually consisting of plutonium blended with natural uranium, reprocessed uranium, or depleted uranium.  MOX fuel is an alternative to the LEU fuel used in light water reactors.
\item[MWth] stands for {\it Megawatts Thermal}.  It is a unit for the total thermal power of a reactor, as opposed to the electrical power, which is typically about one third of the thermal power. 
\item[National Technical Means] are a state's suite of technological capabilities for verification of adherence to treaties. Examples include satellites, radar, and electronic communications systems.
\item[NDA] stands for {\it Non-Destructive Assay}.
\item[NPT] is the {\it Treaty on the Non-Proliferation of Nuclear Weapons}, commonly known as the Non-Proliferation Treaty, an international treaty whose objective is to prevent the spread of nuclear weapons and weapons technology, to promote cooperation in the peaceful uses of nuclear energy, and to further the goal of achieving nuclear disarmament and general and complete disarmament.  It opened for signatures in 1968, and entered into force in 1970.  Currently, 191 states are parties to the NPT.
\item[NNSA] stands for the {\it National Nuclear Security Administration}.
\item[NRC] is the U.S.\ {\it Nuclear Regulatory Commission}, the agency with regulatory authority over all non-defense nuclear activities in the U.S.
\item[Nu Tools] refers to the study that led to this report.  It includes a play on words in which ``Nu" is the English spelling for the Greek letter $\nu$, which is the symbol scientists use for neutrinos, and is pronounced as ``new". 
\item[Overburden] refers to the material used to shield a detector from cosmic rays.  For underground detectors this overburden is the rocks and dirt between surface and the detector.  Overburden is often measured in meters of water-equivalent shielding or (m.w.e.). 
\item[PMDA] is the {\it Plutonium Management and Disposition Agreement}, an agreement between the United States and Russia signed in 2000. It regulates the conversion of non-essential plutonium into mixed oxide (MOX) fuel used to produce electricity. Both sides were required to render a significant amount of their of weapons grade plutonium into reactor grade plutonium alongside reaching a standard for spent fuel to be mixed with more highly irradiating products.
\item[Safeguards-by-Design (SBD)] refers to the consideration of safeguards implementation early in facility planning from design through operation. SBD approaches seek to improve safeguard-ability in a facility's design and reduce safeguards implementation costs at the facility. More details can be found in a safeguards-by-design guidance report\autocite{sbd}. 
\item[Scintillators] are materials that emit optical photons (\textit{scintillation light}) as charged particles lose energy traversing them. \textit{Organic scintillators} comprised of hydrocarbons are commonly used for neutrino detection via the IBD reaction since they provide both the required protons and the means to detect the resulting reaction products.
\item[Significant quantity (SQ)] is the IAEA definition of the approximate amount of nuclear material required for fabrication of a nuclear explosive device.  
\item[SMRs] {\it Small Modular Reactors} are a class of advanced commercial reactor that operate in the 10s-100s of MW range. Many SMR facility designs offer the capability to incorporate multiple reactors, called modules, on a single site. Advantages of SMRs include: lower capital investment, small physical footprint, and flexible power additions with the modules.
\item[TRL] stands for {\it Technical Readiness Level}. The TRL levels of projects or technologies are described is the DOE Technology Readiness Assessment Guide: \url{https://www.directives.doe.gov/directives-documents/400-series/0413.3-EGuide-04-admchg1}.

\end{description}

%% file: interviewees.tex

The following experts were interviewed by members of the Nu Tools Executive Group over the course of this study. These individuals provided views that informed the report, but they were not directly involved in its writing. The Executive Group greatly appreciates the following people for helping inform the study findings and recommendations.

\begin{itemize}

 \TabPositions{5cm} 

\item[]
\item[] Abdalla Abou-Jaoude \tab \textit{Idaho National Laboratory}
\item[] Darius Ancius \tab \textit{Directorate General for Energy, European Commission}
\item[] Jesse Bland \tab \textit{Sandia National Laboratories}
\item[] Mat Budsworth \tab \textit{Atomic Weapons Establishment, United Kingdom}
\item[] Jeff Chapman \tab \textit{NA-84 Nuclear Incident Response, NNSA/ORNL}
\item[] David Chichester \tab \textit{Idaho National Laboratory}
\item[] Bernadette K. Cogswell \tab \textit{Virginia Tech}
\item[] Ferenc Dalnoki-Veress \tab \textit{James Martin Center for Nonproliferation Studies,} \tab \tab \textit{Middlebury Institute of International Studies at Monterey}
\item[] Kevin Deyette \tab \textit{NuScale}
\item[] Mona Dreicer \tab \textit{Lawrence Livermore National Laboratory}
\item[] Andreas Enqvist \tab \textit{University of Florida}
\item[] Rod Ewing \tab \textit{Stanford University}
\item[] Muriel Fallot \tab \textit{Subatech Laboratory (Université de Nantes, CNRS/in2p3,} \tab \tab \textit{IMT Atlantique), France}
\item[] Robert Finch \tab \textit{Sandia National Laboratories}
\item[] George Flanagan \tab \textit{Oak Ridge National Laboratory}
\item[] Alexander Glaser \tab \textit{Princeton University}
\item[] Mark Goodman \tab \textit{U.S. Department of State}
\item[] Bernd Grambow \tab \textit{Subatech, France}
\item[] Siegfried Hecker \tab \textit{Center for International Security and Cooperation, Stanford University}
\item[] Olli Heinonen \tab \textit{Stimson Center}
\item[] James Henkel \tab \textit{NNSA Office of Nuclear Verification}
\item[] David Holcomb \tab \textit{Oak Ridge National Laboratory}
\item[] Michael Hornish \tab \textit{NA-84 Nuclear Search Program, NNSA}
\item[] Allison Macfarlane \tab \textit{University of British Columbia}
\item[] Matthew Malek \tab \textit{University of Sheffield, United Kingdom}
\item[] Christopher Mauger \tab \textit{University of Pennsylvania}
\item[] Vladimir Mozin \tab \textit{Lawrence Livermore National Laboratory}
\item[] Frank Pabian \tab \textit{Stanford University/CISAC Affiliate}
\item[] Todd Palmer \tab \textit{Oregon State University}
\item[] Per Peterson \tab \textit{University of California, Berkeley; Kairos Power}
\item[] David Reyna \tab \textit{Sandia National Laboratories}
\item[] Mark Schanfein \tab \textit{Idaho National Laboratory}
\item[] Pavel Tsvetkov \tab \textit{Texas A\&M University}
\item[] Antonin Vacheret \tab \textit{Imperial College, United Kingdom}
\item[] Klaas van der Meer \tab \textit{SCK CEN, Belgium}
\item[] Louise Evans \tab \textit{Oak Ridge National Laboratory}
\item[] Mital Zalavadia \tab \textit{Pacific Northwest National Laboratory}
\end{itemize}

%% file: miniWorkshop.tex
The purpose of the Nu Tools Mini-Workshop was to engage with the reactor neutrino detector development community.  Although our study is focused on the utility of these technologies and not on the technologies themselves, we believe that it is important to engage with the neutrino scientific community to ensure that we understood their perspective, and that we were aware of any prior work that was done by this community to understand potential uses of their technology.  We were particularly interested in any previous engagement with potential users.  To ensure that this community was aware of our activities, each day's session opened with an overview talk on the Nu Tools study.

The Mini-Workshop consisted of two days of virtual presentations from groups around the world involved in reactor neutrino detector development. Presenters were asked to focus on the applications of their technology and interactions with potential end-users.  Of the 30 groups 
that were invited to participate, 21 chose to give a presentation (see agenda below).  Of these 14 were on inverse beta decay detection and seven were on coherent elastic neutrino nucleus scattering.  The presentations included speakers from ten different countries.  Many of the invited groups that did not choose to present still participated in the workshop discussions. A total of 131 individuals from 14 countries registered to attend.  

\vspace{3mm}
\noindent Presentation slides from the Mini-Workshop can be found at \hyperlink{https://indico.phys.vt.edu/event/43/}{https://indico.phys.vt.edu/event/43/}

\section*{Agenda}
\begin{tabular}{rlll} 
\multicolumn{3}{l}{\textbf{Day One -- Wednesday, July 22, 2020}} \\
Time  & Title & Presenter  \\ \hline
10:00 & Welcome &  \\
10:05 & Nu Tools Overview     & Michael Foxe & \textit{PNNL} \\
      &                       & Nathaniel Bowden & \textit{LLNL} \\
10:30 & PANDA                 & Tomoyuki Konno & \textit{Kitasato University, Japan} \\
11:40 & Ocean Bottom Detector & Hiroko Wantanabe & \textit{Tohoku University, Japan} \\
11:50 & LiquidO               & Pedro Ochoa-Ricoux & \textit{UC Irvine} \\
12:00 & JUNO TAO              & Liang Zhan & \textit{IHEP Beijing, China} \\
12:20 & Efforts in Turkey     & Emrah Tiras & \textit{Iowa State University} \\
12:30 & break                 & \\
13:00 & VIDARR                & Jon Coleman & \textit{University of Liverpool, United Kingdom} \\
13:10 & CHANDLER              & Jonathan Link & \textit{Virginia Tech} \\
13:20 & PROSPECT              & Thomas Langford & \textit{Yale} \\
13:30 & SANDD                 & Steven Dazeley & \textit{LLNL} \\
13:40 & Watchman              & Adam Bernstein & \textit{LLNL} \\
14:00 & ISMRAN                & Lalit Pant & \textit{Bhabha Atomic Research Centre, India} \\
\end{tabular}

\vspace{3mm}
\noindent
\begin{tabular}{rlll} 
\multicolumn{3}{l}{\textbf{Day Two -- Friday, July 24, 2020}} \\
Time  & Title & Presenter  \\ \hline
10:00 & Welcome &  \\
10:05 & Nu Tools Overview     & Jason Newby & \textit{ORNL} \\
      &                       & Nathaniel Bowden & \textit{LLNL} \\
10:30 & CONUS                 & Manfred Lindner & \textit{MPIK, Heidelberg , Germany} \\
10:40 & NUCLEUS               & Raimund Strauss & \textit{TU Munich, Germany} \\
10:50 & Efforts at U. Chicago & Juan Collar & \textit{University of Chicago} \\
11:00 & MINER                 & Rupak Mahapatra & \textit{Texas A\&M} \\
11:10 & RICOCHET              & Steven Weber & \textit{MIT} \\
11:20 & break                 &  \\
11:40 & Nucifer               & Thierry Lasserre & \textit{CEA Saclay, France} \\
11:50 & Angra/CONNIE          & Pietro Chimenti & \textit{Universidade Estadual de Londrina, Brazil} \\
12:00 & vIOLETA               & Ivan Sidelnik & \textit{Comisión Nacional de Energía Atómica, Argentina} \\
12:10 & NuLAT                 & Bruce Vogelaar & \textit{Virginia Tech} \\
12:20 & NUDAR                 & Glenn Jocher & \textit{Ultralytics LLC} \\ 
\end{tabular}